\documentclass[a4paper,9pt]{extarticle}
\usepackage{titling}
\usepackage[utf8]{inputenc}
\usepackage[margin=2.85cm]{geometry} 
\usepackage[hyperfootnotes=false]{hyperref}
\usepackage{indentfirst}
\usepackage{url}

\usepackage{apacite}
\usepackage[round]{natbib}

\usepackage{amsmath}

\usepackage{float} 

\renewcommand{\url}[1]{} 

\usepackage{graphicx}

\DeclareGraphicsExtensions{.pdf,.eps,.tiff,.png,.jpg, .svg}
\usepackage{setspace}
\usepackage{dcolumn}
\usepackage{wasysym}
\usepackage{array}

\usepackage{doi}
\usepackage{authblk}

\renewcommand{\doi}[1]{}

\hypersetup{
    bookmarksopen=false,     
    unicode=false,           
    pdftoolbar=true,         
    pdfmenubar=true,         
    pdffitwindow=false,      
    pdfstartview={FitH},     
    pdfnewwindow=true,       
    colorlinks=true,         
    linkcolor=blue,          
    citecolor=blue,          
    filecolor=black,         
    urlcolor=purple            
}

\usepackage{booktabs}
\usepackage{colortbl}
\usepackage{amssymb}
\usepackage{fancyhdr}
 \usepackage{siunitx} 
\usepackage[framemethod=tikz]{mdframed}
\usepackage{xcolor}

\definecolor{lightlavender}{RGB}{245,240,255}

\newmdenv[
  backgroundcolor=lightlavender,
  linecolor=purple!30,
  linewidth=0.5pt,
  leftmargin=0,
  rightmargin=0,
  innerleftmargin=5pt,
  innerrightmargin=5pt,
  innertopmargin=5pt,
  innerbottommargin=5pt
]{methodbox}

\usepackage{subcaption}

\title{\huge Party Ideologies and Political Polarization-Driven Conflicts: A Study of the Global South}

\author[]{\textbf{Shreyansh Padarha}\thanks{Corresponding author:\href{mailto:shreyansh.padarha@outlook.com}{shreyansh.padarha@outlook.com} \\The code is available at: \href{https://github.com/shreyansh-2003/Party-Ideologies-and-Political-Polarization-Driven-Conflicts-Study}{GitHub Repository}}}

\affil[]{Oxford Internet Institute, University of Oxford, Oxford, UK}

\date{}

\begin{document}

\maketitle

\raggedbottom

\newcommand{\appendixref}[1]{%
  \hyperref[#1]{Appendix~\ref*{#1}}%
}

\newcommand{\figref}[1]{%
  \mbox{\hyperref[#1]{Figure~\ref*{#1}}}%
}

\newcommand{\tabref}[1]{%
  \hyperref[#1]{Table~\ref*{#1}}%
}
\addtolength{\skip\footins}{1em}



\vspace{-3.5em}

\begin{abstract}
\noindent Post-World War II armed conflicts have often been viewed with higher scrutiny in order to avoid a full-scale global war. This scrutiny has led to the establishment of determinants of war such as poverty, inequalities, literacy, and many more. There is a gap that exists in probing countries in the Global South for political party fragmentation and examining ideology-driven polarization's effect on armed conflicts. This paper fills this gap by asking the question: How does political identity-induced polarization affect conflicts in the Global South region? Polarization indices are created based on socially relevant issues and party stances from the V-Party Dataset. Along with control variables, they are tested against the response variables conflict frequency and conflict severity created from the UCDP (Uppsala Conflict Data Program). Through Chow's test, Regional Structural Breaks are found between regions when accounting for polarization-conflict dynamics. A multilevel mixed effects modelling approach is used to create region-specific models to find what types of polarization affect conflict in different geographies and their adherence to normative current developments. The paper highlights that vulnerable regions of the world are prone to higher polarization-induced violence. Modelling estimates indicate polarization of party credo on Minority Rights, Rejection of Political Violence, Religious Principles, and Political Pluralism are strong proponents of cultivated violence. In light of these findings, the paper highlights the theoretical parallels and real-world implications. The Global South's inhibitions and slow progress towards development are caused by hindrances from armed conflicts; this paper's results show self-inflicted political instability and fragmentation's influence on these events, making the case for urgency in addressing and building inter-group homogeneity and tolerance.\vspace{1.25em}

\end{abstract}



\section{Introduction}
\noindent Armed conflicts are known to cause systematic damage to health infrastructure \citep{4} while increasing poverty, unemployment, and homelessness \citep{5,6,7}. They also impact the environmental conditions that in turn cause long-term ailments \citep{8,9}. \noindent Through interpretations of Carl von Clausewitz's seminal work ``\textit{Vom Kriege}" (On War), abetment or will to incite violence can not be measured but only estimated \citep{1}. Such estimates suggest there are certain geographical regions that are unduly affected by social and political violence \citep{3}. Hence, researchers have long tried to answer the question: what drives these conflicts? 

\subsection{Polarization and Political Structure}
\label{section11}
\noindent Grievances and inequalities are often touted as strong factors in forming organizations and groups that conduct organized violence \citep{13,14}. The root of incitement, however, comes from prolonged periods of disagreement amongst political hegemonies. \citet{15} frame this aptly as polarization causing alienation of groups which results in intra-group homogeneity and sense of identity while increasing intergroup heterogeneity. When quantified through polarization indices \citep{16,17,18,19}, such group disagreements has been considered a driving force of organized violence such as political conflicts, non-state violence, and wars. Beyond ideology, this polarization can cause both intra-group \citep{20} and inter-group violence \citep{21}. Similarly, ethnic, minority, and religious polarization in subgroups have also emerged as robust predictors for conflicts \citep{22}.

Political order and structure are fundamentally tied to unrest and violence. \citet{23} show that strong autocracies and democracies are less prone to political instability and conflicts. Current indices and their empirical applications in conflict analysis include macro-level units such as income inequality and political accord (amongst civilians) to quantify polarization. A key limitation in these studies is the failure to capture differences between the representative ``organizations of power". These organizations are political parties, often fighting each other in elections and beyond. Their manifestos and campaigns reflect their ideological stances, and once in power, they act as hegemonies that wield change and enforce laws. This is a significant gap, as we know that socio-political party ideology and allegiance shapes an individual's identity \citep{24,25} and ultimately their decisions. To address this gap, our study takes inspiration from \citet{26} and creates multi-group polarization indices for each country, measuring the differences between party identities and stances.

\subsection{Scope and Focus on Global South}
\noindent Our study focuses on the countries in the \textbf{Global South}, a term popularized post-Cold War. Global South is synonymously used with ``periphery", ``developing" and ``third-world" countries \citep{28}. We exclude countries from North America, Europe, Central Asia along with Australia and New Zealand. The reasoning behind focusing on this specific region stems from the lack of epistemological studies focusing on political-induced polarization in the Global South. Additionally, these countries are more prone to conflicts. Developing countries, especially in Africa, have higher heterogeneity in their populous demographics. These countries see conflicts due to changing religious sentiments, intolerance, and cultural lines \citep{29,30}. Unfortunately, the conflicts invariably affect the marginalized parts of society and women the most \citep{31}, with conflicts in Sub-Saharan Africa increasing cases of domestic sexual violence \citep{32}. Due to the high violence and electoral volatility seen in Latin America \citep{33} and Africa \citep{34}, our election-ideology focused variables are more likely to show inductive relationships in the Global South. 

\subsection{Research Questions}
\noindent In this paper, we try to establish how political fragmentation on party-lines\footnote{The Party Line: political party's position, official ideas and goals \citep{partyline}} within a country affects armed conflict intensity (event rate) and conflict severity (death rate). The hypothesis and underlying research questions mentioned below will be examined within our paper.

\begin{list}{}{\leftmargin=2em \rightmargin=2em}
\item[] \textbf{Hypothesis 1 :} People establish themselves and their identities with affiliated political party ideologies. The differences in polarization of these ideologies result in significant ($+/-$) changes in conflict events rate and conflict-related deaths.
\item[] \textbf{Hypothesis 2 :} Each geographically separate region has different cultural, ethnic, and historical influences. Therefore, the conflict rate (intensity and severity) of regions are caused by significantly different factors (polarization and controls), adhering to normative issues.

\end{list}

\subsection{Related Work and Contributions}
\noindent Most literature in conflict analysis has focused on predicting wars. The most common technique used are Hidden Markov Models (HMMs). HMMs have been used to forecast conflict intensity in the Balkans \citep{37}, likelihood of conflict between Israel-Palestine \citep{38}, and modelling territorial control based in Latin America and Africa \citep{39, 40}. However, HMMs often oversimplify complex systems by using simple state transitions (conflict to no conflict), leading to inaccurate predictions \citep{41}. They require substantial stable data following the Markovian property, where future states depend solely on current states. Despite hybrid regression-Markovian models \citep{42}, limitations persist in interpreting hidden states like the lack of causal inferences and challenges in incorporating multi-dimensional variables.

As mentioned in Section \ref{section11}, studies assessing conflict causation typically focus on macro-level factors. Research on ideological and social polarization tend to be either observational, such as studying Marxist-Leninist ideologies' impact on violence in Mozambique and Angola \citep{43}, or empirical using demographic-division based indices. The latter rely on datasets like Ethnic Power Relations (EPR) \citep{47} to calculate socio-ethnic polarization metrics for conducting conflict modelling \citep{44,45,46}. Similarly, the World Christian Encyclopedia (WEC) \citep{48} and its updates have been considered strong predictors of religious polarization for conflict studies \citep{22,50}. As mentioned above, while these metrics and their empirical conflict modelling are robust in capturing diversity and demographic differences, they fail to represent the ideological differences between the dominant political forces (parties) that ultimately manage, control, or instigate violent conflicts.


 
\section{Methods}
\subsection{Data}
\subsubsection{Conflicts Dataset}
\begin{figure}[b!]
    \centering
    \includegraphics[width=\linewidth]{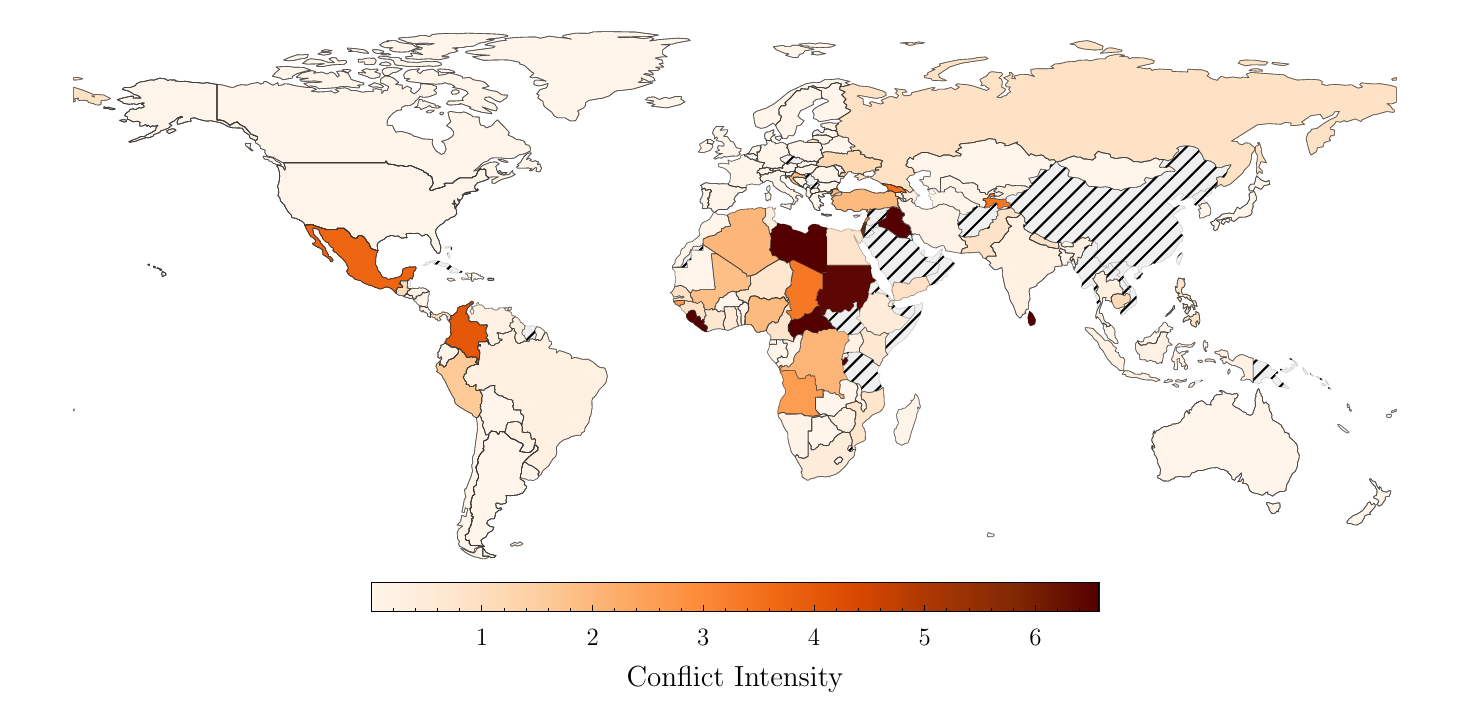}
    \caption{\textbf{World Map of Spread of Conflict Related Death Rate.}
    A visual depiction of the global distribution of conflict-related deaths (1989-2023), normalized per 100,000 population and regime duration (Data from \citet{51}). The map reveals substantially higher conflict intensity in the Global South, particularly across Africa and parts of Latin America, with notably elevated rates in regions experiencing prolonged political instability.}
    \label{mapconflicts}
\end{figure}

\noindent For the response variable (conflicts), we use data from the Uppsala Conflict Data Program (UCDP) \citep{51, 52}. The UCDP georeferenced dataset covers yearly conflict history (1989-2023) for more than 190 countries. The data, collected from multilingual news coverage, encompasses 3 variables of organized violence (state-based, non-state, and one-sided violence). From these, we compute two response variables: \textit{conflict\textsubscript{event-rate}} and \textit{conflict\textsubscript{death-rate}} representing frequency (intensity) and fatality (severity) respectively. These variables are aggregated over each consecutive political regime duration \citep{53}, scaled to 100,000 per capita, and normalized by the regime's time period. As evident in \figref{mapconflicts}, the average conflict death rate has been notably higher in the Global South compared to the North over the past 30 years and has increased dramatically (\appendixref{Appendixconflictrend}).

\subsubsection{Political Identities and Regimes Dataset}
\noindent To quantify heterogeneity among political parties, we utilize the V-Party Dataset \citep{53}, which comprises of expert-annotated political stances expressed by parties during election years. It covers over 1,900 parties across 168 countries. We consider 15 party identities including domains such as social equality (minority rights, gender equality, working women), economic ideologies (left-right scale, welfare), and religious principles, described in \appendixref{tableVars}. The dataset scores political identities on 0-4, 0-5, or 0-6 scales, with endpoints representing opposites like oppose-support or left-right. The numerical mapping of responses \citep{54} is redundant when creating polarization metrics, as we treat each identity as an independent spectrum. We consider 94 countries for our study categorized into 4 different regions Middle East and North Africa (MENA), Sub-Saharan Africa (SSA), Latin America and the Caribbean (LAC), Asia and Pacific (AP). The time period of our study is condensed between 1989 to 2023. An \textit{era} categorical variable is created with Jenks Natural Break Optimization \citep{jenks}, creating 4 time periods ($1989-2000, \ 2001-2008, \ 2009-2015, \ 2016-2023$).

\subsubsection{Readjusted Dalton’s Polarization Index}
\noindent We adjust Dalton's formula \citep{55} of left-right polarization spectrum to create our Dalton's Readjusted Index (DRI). We use party seat share rather than vote share as weights, given the prevalence of minority governments and coalition formations \citep{56}. This follows Gamson's law \citep{57}, which determines power distribution in governments by finding that cabinet portfolios (policy influence) are allocated proportionally to each party's seat contribution in the coalition \citep{58}. Additionally, electoral systems' vote-to-seat conversion mechanisms significantly impact parties' legislative effectiveness \citep{59, 60, 61}, making seat share a more direct measure of political influence. For each country's unique election period and political ideology (identity) considered, we calculate the polarization as

\begin{equation}
\text{Polarization \textsubscript{$(DRI)$}} = \sqrt{\sum{i=1}^{n} s_i \left(\frac{p_i-\bar{p}}{R/2}\right)^2}
\end{equation}

where for each party $i$, $s_i$ is its seat share, $p_i$ is its political stance score for a given identity, $\bar{p} = \sum_{i=1}^{n} s_i p_i$ is the weighted mean of political stance scores across all parties, $R$ is the response (stance) scale range, and $n$ is the number of parties in the system.

\subsubsection{Control Variables}
\noindent The selected controls for the study include population size, regime length, GDP per capita, Gini coefficient, religious freedom and freedom of expression indices. A 1\% GDP increase reduces civil war probability by 0.5 \citep{10, 63}, while income inequality measured through the Gini coefficient consistently correlates with increased political unrest \citep{64, 65, 66, 67}. Similarly, religious freedom serves as a crucial control, as it can both mitigate conflict through protected practices \citep{29} and heighten tensions when intersecting with political mobilization \citep{70, 71}, with religious discrimination increasing conflict odds 2.6 fold \citep{29}. Likewise, freedom of expression demonstrates conflict-reducing properties in democracies \citep{72} and decreases terrorism likelihood \citep{73}. We use the V-Dem \citep{53} dataset for the variables population size, religious freedom, and freedom of expression. The World Bank Repository is used for finding the parity adjusted GDP per capita and Gini coefficient. All variables are converted from year-wise data to average values for each regime, based on its time period.


\subsection{Regional Structural Breaks with Chow's Test}
\noindent Chow's test is a statistical technique used for identifying structural breaks within data \citep{74}, helping evaluate whether two subsets can be pooled together. We divide the data according to different regions and fit three OLS (Ordinary Least Squares) models: two separate models for each region pair and one pooled model combining both regions. For each region pair $(r_1, r_2)$, we estimate

\begin{equation} \label{eq:region1}
y_{r1} = \alpha_1 + \underbrace{\sum_{i=1}^{15} \beta_{1i}P_i}_{\text{Polarization \ Variables}} + \underbrace{\gamma_1G + \delta_1Y + \theta_1N + \lambda_1R + \phi_1F}_{\text{Controls}} + \varepsilon_1
\end{equation}

\begin{equation} \label{eq:region2}
y_{r2} = \alpha_2 + \underbrace{\sum_{i=1}^{15} \beta_{2i}P_i}_{\text{Polarization \ Variables}} + \underbrace{ \gamma_2G + \delta_2Y + \theta_2N + \lambda_2R + \phi_2F}_{\text{Controls}} + \varepsilon_2
\end{equation}

\begin{equation} \label{eq:pooled}
y = \alpha + \underbrace{\sum_{i=1}^{15} \beta_{i}P_i}_{\text{Polarization \ Variables}} + \underbrace{\gamma G + \delta Y + \theta N + \lambda R + \phi F}_{\text{Controls}} + \varepsilon
\end{equation}

where $y$ denotes the response variables \textit{conflict\textsubscript{event-rate}} / \textit{conflict\textsubscript{death-rate}}, $P_i$ represents polarization variables, $G$ is Gini coefficient, $Y$ is GDP per capita, $N$ is population, $R$ and $F$ are religious freedom and freedom of expression indices respectively. The Chow test statistic is

\begin{equation} \label{eq:chow}
F = \frac{(RSS_p-RSS_{r1}-RSS_{r2})/k}{(RSS_{r1} + RSS_{r2})/(n_1 + n_2-2k)} \sim F_{k, n_1 + n_2-2k}
\end{equation}

where $RSS_p$ is the residual sum of squares from equation \ref{eq:pooled}, $RSS_{r1}$ and $RSS_{r2}$ are from equations \ref{eq:region1} and \ref{eq:region2} respectively, $k$ is the number of parameters, and $n_1, n_2$ are regional sample sizes.  If the Chow F-statistic is found to be significant, there exists structural break and deep differences between model coefficient and regional data.

\subsection{Modelling Strategy}

\begin{figure}[H]
    \centering
    \begin{subfigure}[b]{0.44\textwidth}
        \centering
        \includegraphics[width=\textwidth]{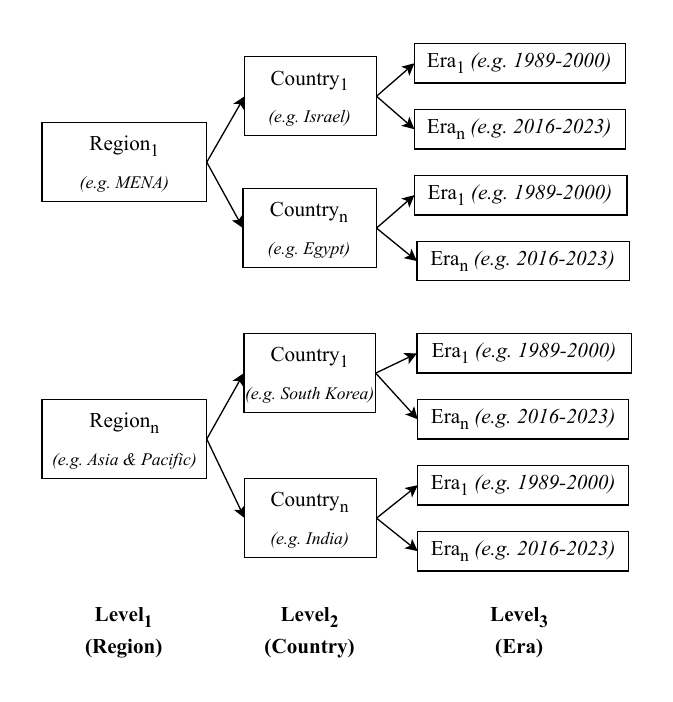}
        \caption{Baseline Model}
        \label{baselinearch}
    \end{subfigure}
    \hfill
    \begin{subfigure}[b]{0.44\textwidth}
        \centering
        \includegraphics[width=\textwidth]{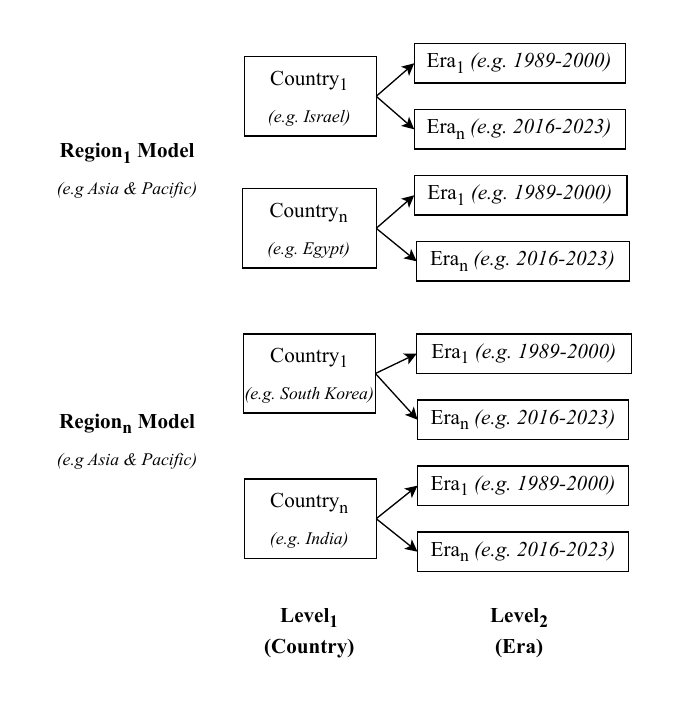}
        \caption{Region-wise Separate Models}
        \label{regionwisearch}
    \end{subfigure}
    \caption{\textbf{Mixed Effects Multilevel Modelling Structure.} \textbf{(a)} The baseline model has three nested levels: Region (N\textsubscript{groups} = 4), Countries (N\textsubscript{groups} vary), Era's (N\textsubscript{groups} = 5, barring some countries). \textbf{(b)} The auxiliary modelling strategy we use is to create separate mixed effects models for each region. These individual models take use region's data subset and involve 2 levels: Country and Era.}
    \label{multilevelarch}
\end{figure}
\noindent Mixed effects models are widely used in armed conflict analysis with healthcare \citep{75}. Given our data's multiple groupings with varying but small sample sizes, we employ multilevel (hierarchical) mixed effects modelling rather than fixed effects (\figref{multilevelarch}).

\subsubsection{Baseline Model}
\noindent The baseline model incorporates three levels: region, country, and era (\figref{baselinearch}). We estimate the linear mixed effects models as

\begin{subequations}
\begin{align}
y = \alpha + \overbrace{\underbrace{\sum_{i=1}^{15} \beta_i P_i}_{\text{Polarization Variables}} + 
    \underbrace{\gamma G + \delta Y + \theta N + \lambda R + \phi F}_{\text{Controls}}}^{\text{Fixed Effects}} 
    + \overbrace{\underbrace{b_{\text{region}} + b_{\text{country}} + b_{\text{era}}}_{\text{Random Effect Intercepts}}}^{\text{Random Part}} + \varepsilon \\
b_{\text{region}} \sim N(0,\Sigma_c) \\
b_{\text{country}} \sim N(0,\Sigma_c) \\
b_{\text{era}} \sim N(0,\Sigma_e) \\
\varepsilon_r \sim N(0,\sigma^2)
\end{align}
\end{subequations}

where the polarization variables ($P_i$) and controls ($G$, $Y$, $N$, $R$ and $F$) are the same as in equation \ref{eq:pooled} and $y$ takes the shape of either \textit{conflict\textsubscript{event-rate}} or \textit{conflict\textsubscript{death-rate}}. The random part contains $b_{\text{region}}$, $b_{\text{country}}$, and $b_{\text{era}}$ which represent random intercepts for each region, country, and era respectively. $\Sigma$ captures the variance components of these random effects and $\varepsilon$ is the observation-level error with variance $\sigma^2$. They are clustered and estimated with era-country groupings through a nested dummy variable.


\subsubsection{Region-wise Modelling}
\noindent Based on Chow's test for structural analysis (\ref{sectionChow}) and group-level variance (region and country) observed in the baseline model (\tabref{tableBaseline}), we develop separate mixed effects models for each region using country and era as the two levels (\figref{regionwisearch}). The conflict variables are estimated for each region ($r$) as

\begin{subequations}
\begin{align}
y_r = \alpha + \overbrace{\underbrace{\sum_{i=1}^{15} \beta_i P_i}_{\text{Polarization Variables}} + 
    \underbrace{\gamma G + \delta Y + \theta N + \lambda R + \phi F}_{\text{Controls}}}^{\text{Fixed Effects}} 
    + \overbrace{b_{\text{country}} + b_{\text{era}}}^{\text{Random Part}} + \varepsilon_r\\
b_{\text{country}} \sim N(0,\Sigma_c) \\
b_{\text{era}} \sim N(0,\Sigma_e) \\
\varepsilon_r \sim N(0,\sigma^2)
\end{align}
\end{subequations}

where the polarization variables ($P_i$) and controls ($G$, $Y$, $N$, $R$ and $F$) are the same as in equation \ref{eq:pooled}, with $y_r$ representing the conflict rates for region $r$. The random part now contains only $b_{\text{country}}$ and $b_{\text{era}}$ which represent random intercepts for each country and era within the region respectively. Once the model is fitted, we test the residuals and coefficients against standard assumptions and conduct validation checks for robustness (\ref{sectionVALIDATE}).


\newpage

\section{Results}

\subsection{Regional Structure Break Analysis}
\label{sectionChow}
\begin{figure}[h]
    \centering
    \includegraphics[width=0.65\linewidth]{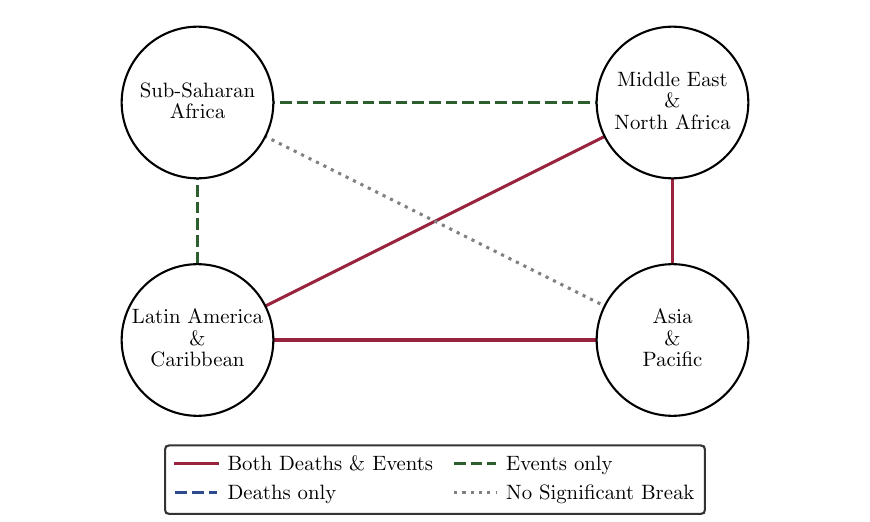}
\caption{\textbf{Network Visualization of Regional Structure Breaks in Conflict-Polarization Dynamics.} The Network shows structural breaks across regions based on Chow's tests. Solid burgundy lines show breaks in both death and event rates, dashed lines show breaks in death (blue) or event rates (green). Sub-Saharan Africa and Asia \& Pacific show most similar coefficients with no breaks (dotted gray line).}
    \label{structuralbreaksnetwork}
\end{figure}

\begin{figure}[h]
    \centering
    \includegraphics[width=0.61\linewidth]{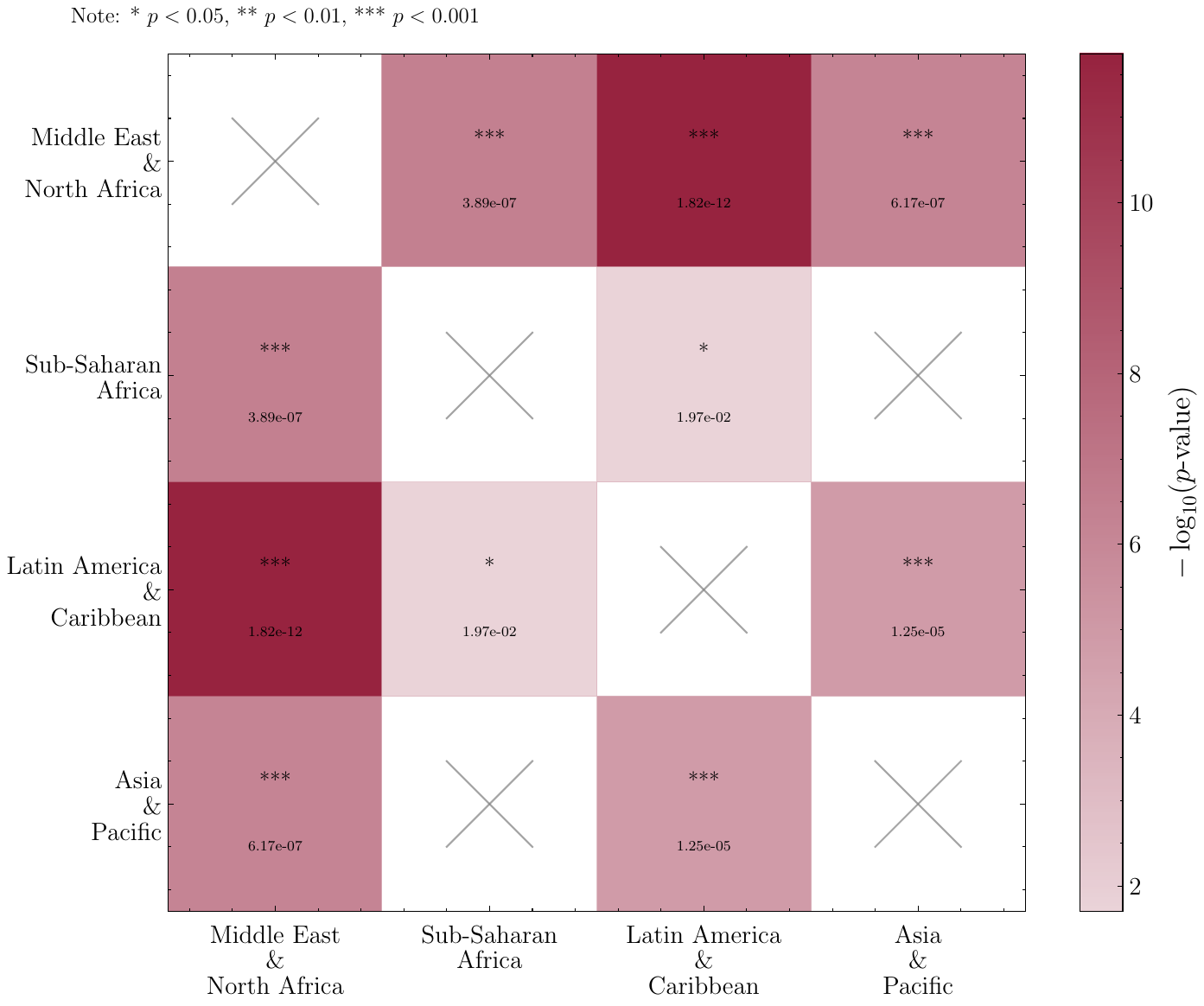}
    \caption{\textbf{Heatmap of Chow's Regional Heterogeneity Test.} The figure visualises the pairwise Chow's test results using $-\log_{10}$ transformed p-values, where darker burgundy indicates stronger statistical significance. Cross marks (×) represent non-significant differences ($p > 0.05$). The Middle East \& North Africa exhibits the strongest structural breaks with other regions.}.
    \label{heatmapchows}
\end{figure}

\noindent Using Chow tests to examine regional heterogeneity in the relationship between polarization (with controls) and conflict variables, we observe five out of six regional pairs exhibit significant differences in OLS coefficients (\figref{structuralbreaksnetwork}) for both \textit{conflict\textsubscript{death-rate}} and \textit{conflict\textsubscript{event-rate}}. Sub-Saharan Africa shows the most similar coefficients, with no structural breaks with Asia and Pacific, and differences only in \textit{conflict\textsubscript{event-rate}} with other regions. On $-\log_{10}$ transformation of the p-values, the Middle East \& North Africa region exhibits the most distinct variables across all regional model relationships (\figref{heatmapchows}).

\vspace{1em}

\subsection{Baseline Model}

\begin{table}[b!]
\centering
\caption{\textbf{Baseline Linear Mixed Effects Regression Results.} Presents baseline model estimates for \textit{conflict\textsubscript{event-rate}} and \textit{conflict\textsubscript{death-rate}}. Columns 1-2 show significance with +/- symbols, while columns 3-4 show coefficients. ICC measures variance explained by region, country and era-level grouping. Breusch-Pagan test examines residual heteroskedasticity. Higher \textit{religious principles} and \textit{gender equality} polarization increases conflicts, while lower \textit{freedom of expression} and \textit{LGBT equality polarization} reduces conflict deaths. Both the response variables have high variance in their residuals (heteroskedastic).}
\begin{tabular}{l c c r@{}l r@{}l}
\toprule[1.5pt]
& \multicolumn{2}{c}{Significance} & \multicolumn{4}{c}{Coefficients} \\
\cmidrule(lr){2-3} \cmidrule(lr){4-7}
& Event Rate & Death Rate & \multicolumn{2}{c}{Event Rate} & \multicolumn{2}{c}{Death Rate} \\
\midrule
\multicolumn{7}{l}{\textbf{Polarization Variables}} \\
Anti-Elitism & & & $-$0&.043 & $-$0&.259 \\
People-Centrism & & & 0&.040 & 0&.832 \\
Political Opponents & & & $-$0&.003 & $-$0&.363 \\
Political Pluralism & $--$ & $-$ & $-$0&.628$^{**}$ & $-$1&.995$^{*}$ \\
Minority Rights & & & 0&.082 & 1&.274 \\
Rejection of Political Violence & & & $-$0&.293 & $-$0&.559 \\
Immigration & $+$ & & 0&.368$^{*}$ & 1&.295 \\
LGBT Social Equality & & $--$ & $-$0&.157 & $-$1&.659$^{**}$ \\
Cultural Superiority & & & 0&.230 & $-$0&.259 \\
Religious Principles & $+++$ & $++$ & 0&.491$^{***}$ & 1&.912$^{**}$ \\
Gender Equality & $+++$ & & 0&.580$^{***}$ & 1&.347 \\
Working Women & & & $-$0&.067 & 0&.938 \\
Economic Left-Right Scale & & & 0&.144 & 0&.095 \\
Welfare & & & 0&.007 & $-$0&.318 \\
Clientelism & & & 0&.017 & 0&.447 \\
\midrule
\multicolumn{7}{l}{\textbf{Control Variables}} \\
Average Term Population & $--$ & & $-$0&.000$^{**}$ & $-$0&.000 \\
Period Length & & & 0&.012 & 0&.010 \\
Freedom of Expression & $--$ & $--$ & $-$0&.377$^{**}$ & $-$2&.011$^{**}$ \\
Religious Freedom & & & 0&.033 & 0&.026 \\
GDP per Capita & & & $-$0&.000 & $-$0&.000 \\
Gini Index & & & $-$0&.001 & $-$0&.028 \\
\midrule
\multicolumn{7}{l}{\textbf{Random Effects Parameters}} \\
Constant & & $++$ & 0&.196 & 1&.713$^{**}$ \\
Region Variance & & & 0&.013 & 8&.570 \\
Country Variance & & & 0&.000 & 0&.005 \\
Era Variance & & & 0&.006 & 1&.841 \\
\midrule
\multicolumn{7}{l}{\textbf{Model Summary}} \\
& \multicolumn{2}{c}{Event Rate} & \multicolumn{4}{c}{Death Rate} \\
\cmidrule(lr){2-3} \cmidrule(lr){4-7}
R² Adjusted & \multicolumn{2}{c}{0.241} & \multicolumn{4}{c}{0.647} \\
R² Marginal & \multicolumn{2}{c}{0.268} & \multicolumn{4}{c}{0.660} \\
R² Conditional & \multicolumn{2}{c}{0.732} & \multicolumn{4}{c}{0.925} \\
ICC Region & \multicolumn{2}{c}{0.604} & \multicolumn{4}{c}{0.378} \\
ICC Country & \multicolumn{2}{c}{0.032} & \multicolumn{4}{c}{0.424} \\
ICC Era & \multicolumn{2}{c}{0.000} & \multicolumn{4}{c}{0.000} \\
Heteroskedasticity (p-value) & \multicolumn{2}{c}{0.000$^{***}$} & \multicolumn{4}{c}{0.011$^{**}$} \\
N & \multicolumn{2}{c}{604} & \multicolumn{4}{c}{604} \\
\bottomrule[1.5pt]
\addlinespace[1ex]
\multicolumn{7}{l}{\small\textit{Note:}} \\
\multicolumn{7}{l}{\small +/-/* p $<$ 0.10 , ++/--/** p $<$ 0.05, +++/---/*** p $<$ 0.01} \\
\multicolumn{7}{l}{\small + positive effect, - negative effect} \\
\label{tableBaseline}
\end{tabular}
\end{table}

\noindent The baseline mixed effects model results (\tabref{tableBaseline}) reveal that regional clustering accounts for 60.4\% of variance in \textit{conflict\textsubscript{event-rate}} and 37.8\% in \textit{conflict\textsubscript{death-rate}}, while country-level effects explain 3.2\% and 42.4\% respectively. A significant positive coefficient (+) indicates that as the variable increases, conflict rates increase, while negative coefficients (-) suggest that as the variable increases, conflict rates decrease. There are two variables that are significant ($p<0.01$) across both \textit{conflict\textsubscript{event-rate}} and \textit{conflict\textsubscript{death-rate}}: \textit{Freedom of Expression} (decreases both) and \textit{Religious Principles Polarization} (increases both). The number of polarization variables that achieve significance differs between models, with event rates showing more significant relationships (Immigration, Gender Equality, Religious Principles, all positive) compared to death rates (\textit{LGBT Social Equality} negative, \textit{Religious Principles} positive). Both models demonstrat\textit{}e significant Heteroskedasticity ($p<0.01$ for events, $p<0.05$ for deaths), with higher explanatory power for death rates ($R^2_{conditional}=0.925$) compared to event rates ($R^2_{conditional}=0.732$).

\subsection{Region-wise Models}
\begin{figure}[H]
    \centering
    \hspace{2cm}
    \includegraphics[width=0.83\linewidth]{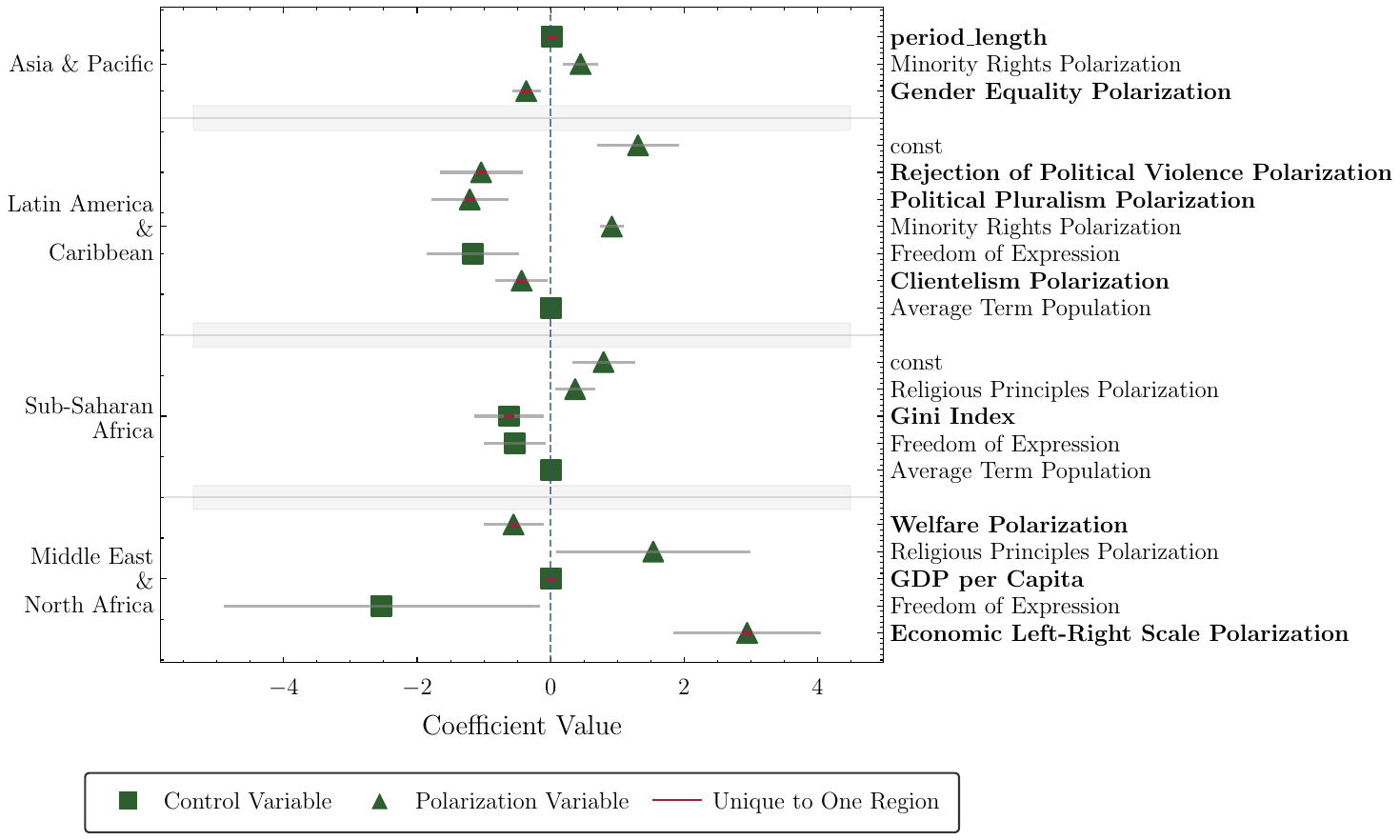}
    \caption{\textbf{Significant Coefficients for Conflict\textsubscript{event-rate} (Intensity).}
    Looking at the significant coefficients for \textit{conflict\textsubscript{event-rate}} across models, the control variables (\textit{Freedom of Expression, Average Term Population}) maintain consistent significance across multiple regions. Region-specific models show distinct patterns. The Middle East and North Africa specific model is uniquely influenced by\textit{ Economic Left-Right Scale Polarization} (+++). Sub-Saharan Africa estimates show strong religious-economic influence (\textit{Religious Principles} +, \textit{Gini Index} -). Asia and Pacific predictors are distinctly affected by identity-based polarization (\textit{Gender Equality} --, \textit{Minority Rights} +). Latin America and Caribbean conflict intensity is characterized by institutional-social tensions (\textit{Clientelism} -, \textit{Minority Rights} +).}
    \label{coeffER}
\end{figure}

\begin{figure}[H]
    \centering
    \hspace{2cm}
    \includegraphics[width=0.83\linewidth]{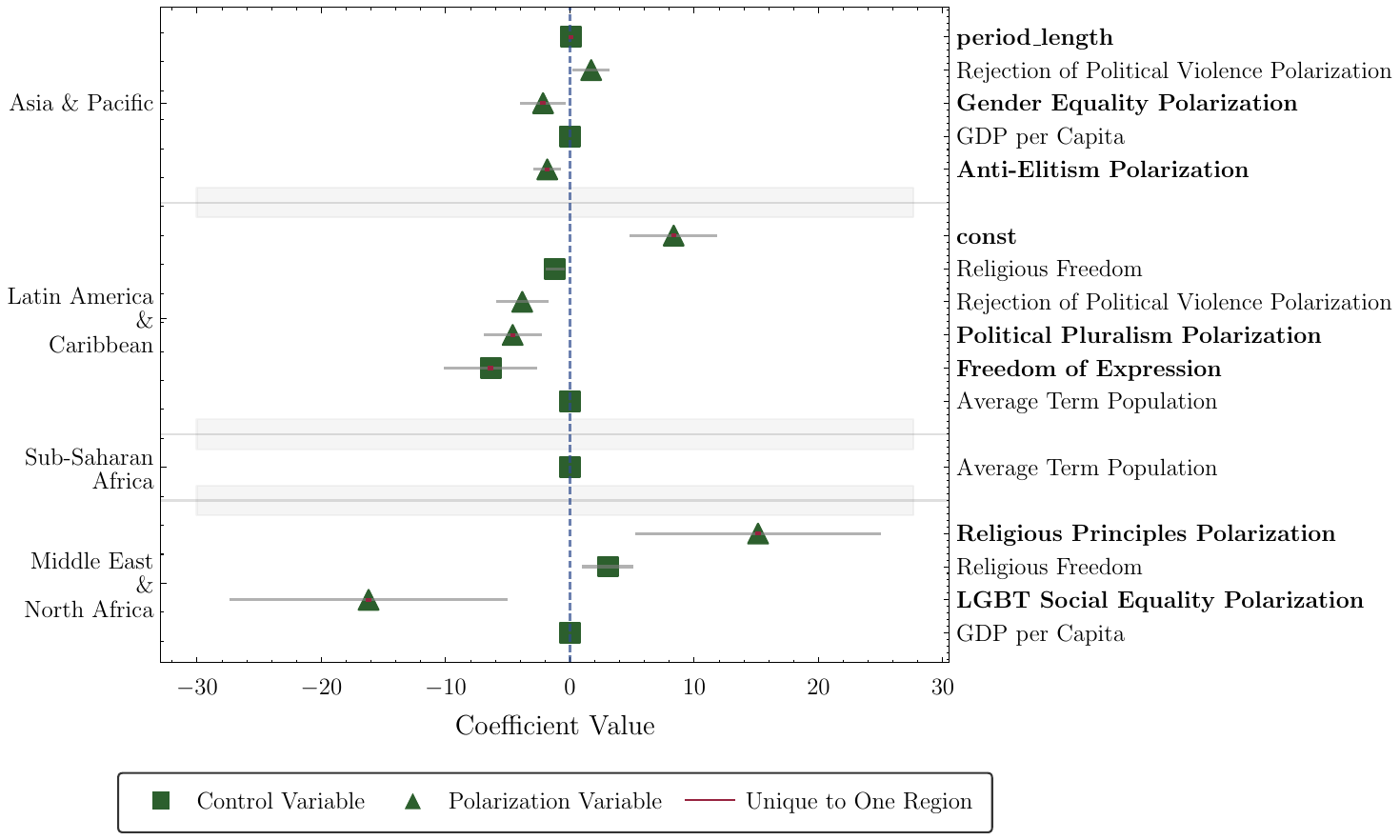}
    \caption{\textbf{Significant Coefficients for Conflict\textsubscript{death-rate} (Intensity).}
    Like conflict intensity models (\figref{coeffER}), freedom controls (\textit{Religious Freedom, Freedom of Expression}) appear consistently across regions. Unlike \textit{conflict\textsubscript{event-rate}}, violence polarization shows positive associations where significant. MENA's death rate is distinctly driven by \textit{LGBT Social Equality Polarization} (---), with strong effects from \textit{Religious Principles Polarization} (+++) and \textit{GDP per Capita} (---). Asia \& Pacific shows strong \textit{Anti-Elitism Polarization} (++) and economic influences. Latin America exhibits interplay of \textit{Political Pluralism} and \textit{Violence Rejection}. Sub-Saharan Africa shows minimal significance, with only \textit{Average Term Population} (+) having an effect on the outcome variable.}
\label{coeffDR}
\end{figure}

\noindent The region-specific models show substantial heterogeneity in both the types of polarization influencing conflicts and their magnitude (\figref{coeffER}, \figref{coeffDR}). The results (\tabref{tableRegion}) reveal that some variables maintain consistent effects across regions, such as \textit{Religious Principles Polarization}, which shows positive associations in MENA and Sub-Saharan Africa. Other variables have region-specific impacts. This is evident in Latin America and the Caribbean, where polarization of \textit{Rejection of Political Violence} has a strong positive effect on both  \textit{conflict\textsubscript{event-rate}} and \textit{conflict\textsubscript{death-rate}}. Control variables show more consistent patterns across regional models compared to polarization variables. \textit{Average Term Population} maintains a significant positive relationship across most regional models, indicating higher population levels are more susceptible to conflicts. Similarly, higher levels of GDP per capita lead to lower conflict deaths and events in MENA and Asia \& Pacific.

\begin{figure}[b!]
   \centering
   \begin{subfigure}[b]{0.49\textwidth}
      \centering
      \includegraphics[width=\textwidth]{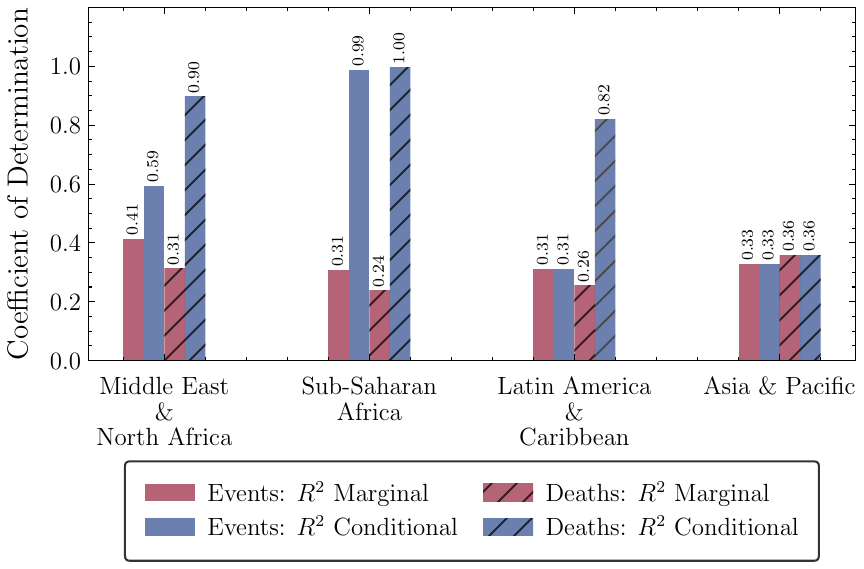}
      \caption{\centering Marginal and Conditional R-Squared ($R^2$) Values Comparison}
      \label{r2}
   \end{subfigure}
   \hfill
   \begin{subfigure}[b]{0.49\textwidth}
      \centering
      \includegraphics[width=\textwidth]{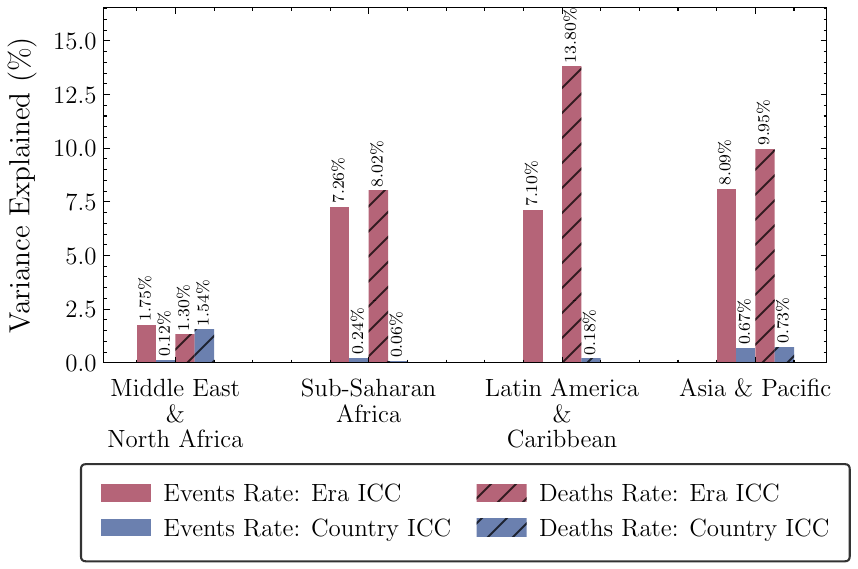}
      \caption{\centering Intraclass correlation coefficients (ICC) Variance Explained Comparison}
      \label{icc}
   \end{subfigure}
\caption{\textbf{Regional Model Performance and Variance Decomposition Analysis.} \textbf{(a)} Comparing R-squared values, Sub-Saharan Africa shows highest performance ($R^2_c \approx 1.0$, $R^2_m \approx 0.3$), while Asia \& Pacific exhibits the least impact with minimal difference between marginal and conditional values. \textbf{(b)} ICC (\(\rho = \frac{\sigma^2_{\text{between}}}{\sigma^2_{\text{between}} + \sigma^2_{\text{within}}} \)) shows variance from country/era clustering in null models. Latin America has highest era-level clustering (13.8\%), while MENA shows minimal clustering (ICC $<$ 2\%) despite strong model fit. Overall ICC values remain low (ICC $<$ 15\%), with particularly weak country-level grouping effects.}
   \label{fig:combined}
\end{figure}

The ICC (Intraclass Correlation Coefficients) represents explanation power of the random effects (country, era) on conflict prediction in a null model (no predictors). We observe low values of conflict response variable explanation in the region-wise models (\figref{icc}). Within the observations, time (era) explains significantly more data variance than country. This pattern changes significantly when the random effects are combined with the fixed effects. There is a profound difference between marginal (variance explained by fixed effects) and conditional \textit{R}$^2$ (variance explained by both fixed and random effects) values, with the latter being higher in most regional models, barring Asia and Pacific (\figref{r2}). This demonstrates that the differences in country and time period within regions significantly impact conflict intensity and severity when considered along with polarization and control variables.

\begin{table}[H]
    \centering

\caption{\textbf{Region-wise Modelling: Linear Mixed Effects Regression Results.} Region-wise models are estimated for \textit{conflict\textsubscript{event-rate}} (ER) and \textit{conflict\textsubscript{death-rate}} (DR) across four regions. Regions are abbreviated by excluding ``The Caribbean" from Latin America and ``Africa" from Sub-Saharan. Most controls (except Gini index and period length) appear in multiple region models. Population correlates positively with conflicts in Sub-Saharan Africa and Latin America, but inversely in MENA. Religious Principles polarization and Religious Freedom show positive relationships with response variables in all regions except Asia-Pacific. Exact coefficients values are shown in \appendixref{modelresultsAppendix}.}
\begin{tabular}{l *{4}{cc}}
\toprule[1.5pt]
& \multicolumn{2}{c}{MENA} & \multicolumn{2}{c}{Sub-Saharan} & \multicolumn{2}{c}{Latin America} & \multicolumn{2}{c}{Asia \& Pacific} \\
\cmidrule(lr){2-3} \cmidrule(lr){4-5} \cmidrule(lr){6-7} \cmidrule(lr){8-9}
& ER & DR & ER & DR & ER & DR & ER & DR \\
\midrule
\multicolumn{9}{l}{\textbf{Polarization Variables}} \\
Anti-Elitism & & & & & & & & $---$ \\
People-Centrism & & & & & & & & \\
Political Opponents & & & & & & & & \\
Political Pluralism & $--$ & & & & $---$ & $---$ & & \\
Minority Rights & & + & & & +++ & & +++ & \\
Rejection of Political Violence &$-$& & & & $---$ & $---$ & + & ++ \\
Immigration & & &$-$& & & & & \\
LGBT Social Equality & & $---$ & & & & & & \\
Cultural Superiority & + & & & & + & & & \\
Religious Principles & ++ & +++ & ++ & + & & & & \\
Gender Equality & & & & & & & $---$ & $--$ \\
Working Women & & & & & & & & \\
Economic Left-Right & +++ & & & & & & & \\
Welfare & $--$ & & & & + & & & \\
Clientelism & & & & & $--$ & & & \\
\midrule
\multicolumn{9}{l}{\textbf{Control Variables}} \\
Avg. Term Population & $--$ & & +++ & ++ & +++ & +++ & & \\
Period Length & & & & & & & +++ & ++ \\
Freedom Expression & $--$ & & $--$ & & $---$ & $---$ & & \\
Religious Freedom & & +++ & & & & $---$ & & \\
GDP per Capita & $--$ & $---$ & & & & & & $---$ \\
Gini Index & & & $--$ & & & & & \\
\midrule
\multicolumn{9}{l}{\textbf{Model Summary}} \\
& \multicolumn{2}{c}{MENA} & \multicolumn{2}{c}{Sub-Saharan} & \multicolumn{2}{c}{Latin America} & \multicolumn{2}{c}{Asia \& Pacific} \\
\cmidrule(lr){2-3} \cmidrule(lr){4-5} \cmidrule(lr){6-7} \cmidrule(lr){8-9}
& ER & DR & ER & DR & ER & DR & ER & DR \\
R² Adjusted & 0.09 & 0.07 & 0.24 & 0.16 & 0.22 & 0.15 & 0.18 & 0.22 \\
R² Marginal & 0.41 & 0.31 & 0.31 & 0.24 & 0.31 & 0.26 & 0.33 & 0.36 \\
R² Conditional & 0.59 & 0.90 & 0.99 & 1.00 & 0.31 & 0.82 & 0.33 & 0.36 \\
Heteroskedasticity & 0.07 & 0.27 & \textcolor{red}{0.00} & \textcolor{red}{0.00} & 0.08 & 0.08 & 0.20 & 0.13 \\
N & \multicolumn{2}{c}{63} & \multicolumn{2}{c}{236} & \multicolumn{2}{c}{178} & \multicolumn{2}{c}{126} \\
\bottomrule[1.5pt]
\addlinespace[1ex]
\multicolumn{9}{l}{\small\textit{Note:}} \\
\multicolumn{9}{l}{\small +/- p $<$ 0.10 , ++/$--$ p $<$ 0.05, +++/$---$ p $<$ 0.01} \\
\multicolumn{9}{l}{\small + positive effect,$-$negative effect} \\
\multicolumn{9}{l}{\small Red values in heteroskedasticity row indicate statistical significance (p $<$ 0.05)} \\
\label{tableRegion}
\end{tabular}
\end{table}

\subsection{Model Assumptions and Validation}
\label{sectionVALIDATE}
\begin{figure}[t!]
    \centering
    \begin{subfigure}[b]{0.49\textwidth}
        \includegraphics[width=\textwidth]{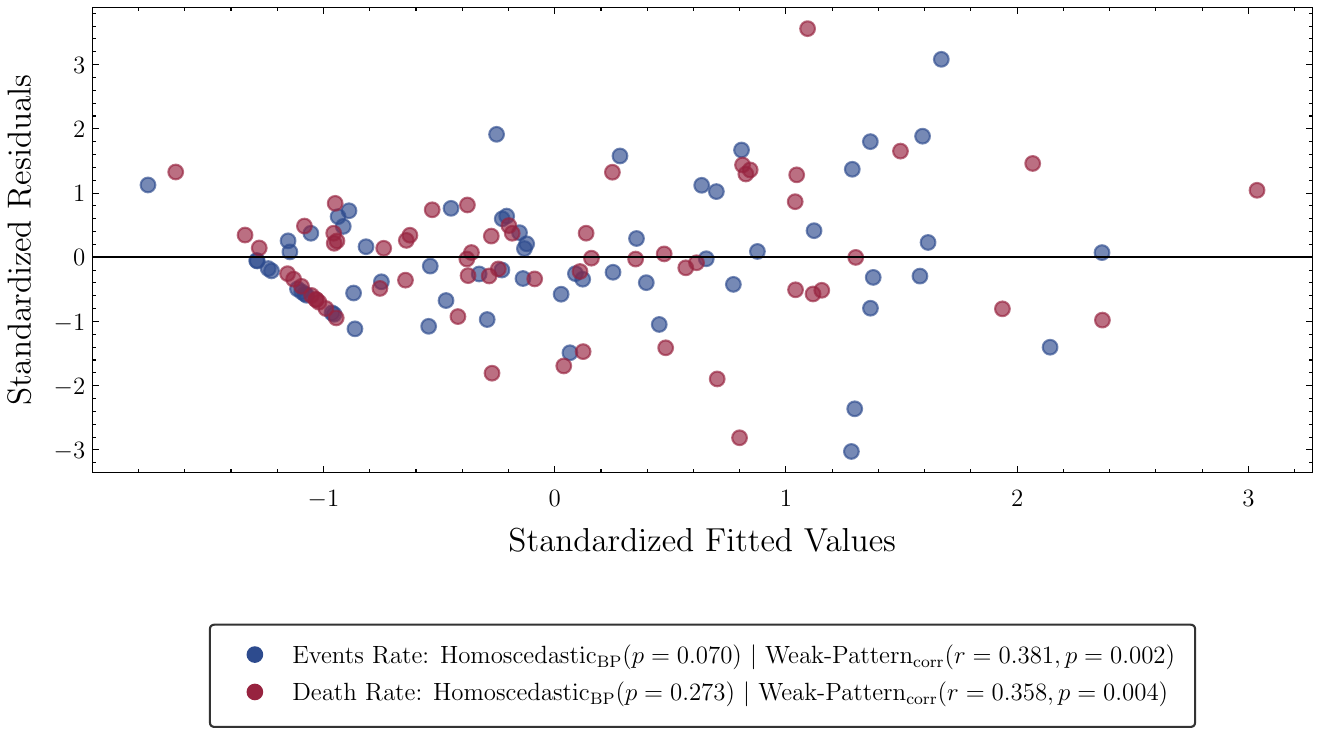}
        \caption{Middle East and North Africa (MENA)}
        \label{heteroMENA}
    \end{subfigure}
    \hfill
    \begin{subfigure}[b]{0.49\textwidth}
        \includegraphics[width=\textwidth]{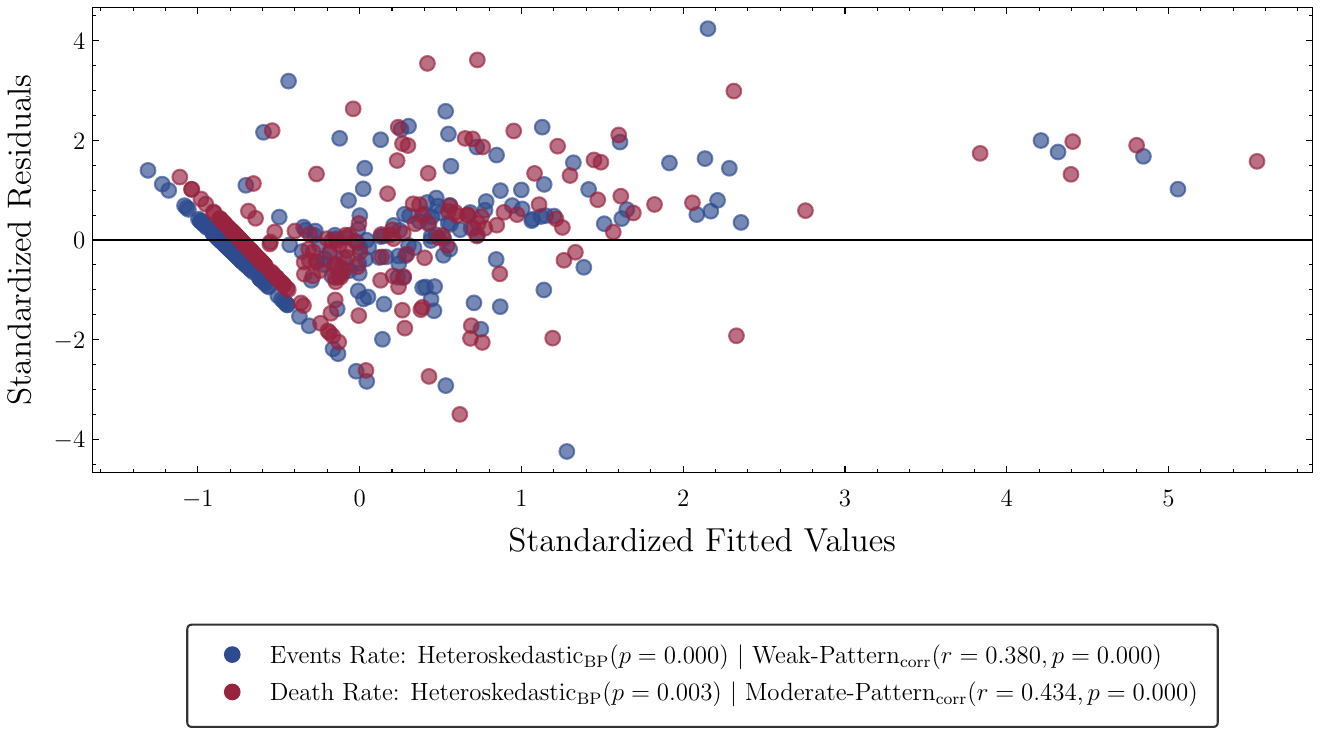}
        \caption{Sub-Saharan Africa}
        \label{heteroSSA}
    \end{subfigure}
    \vskip\baselineskip
    \begin{subfigure}[b]{0.49\textwidth}
        \includegraphics[width=\textwidth]{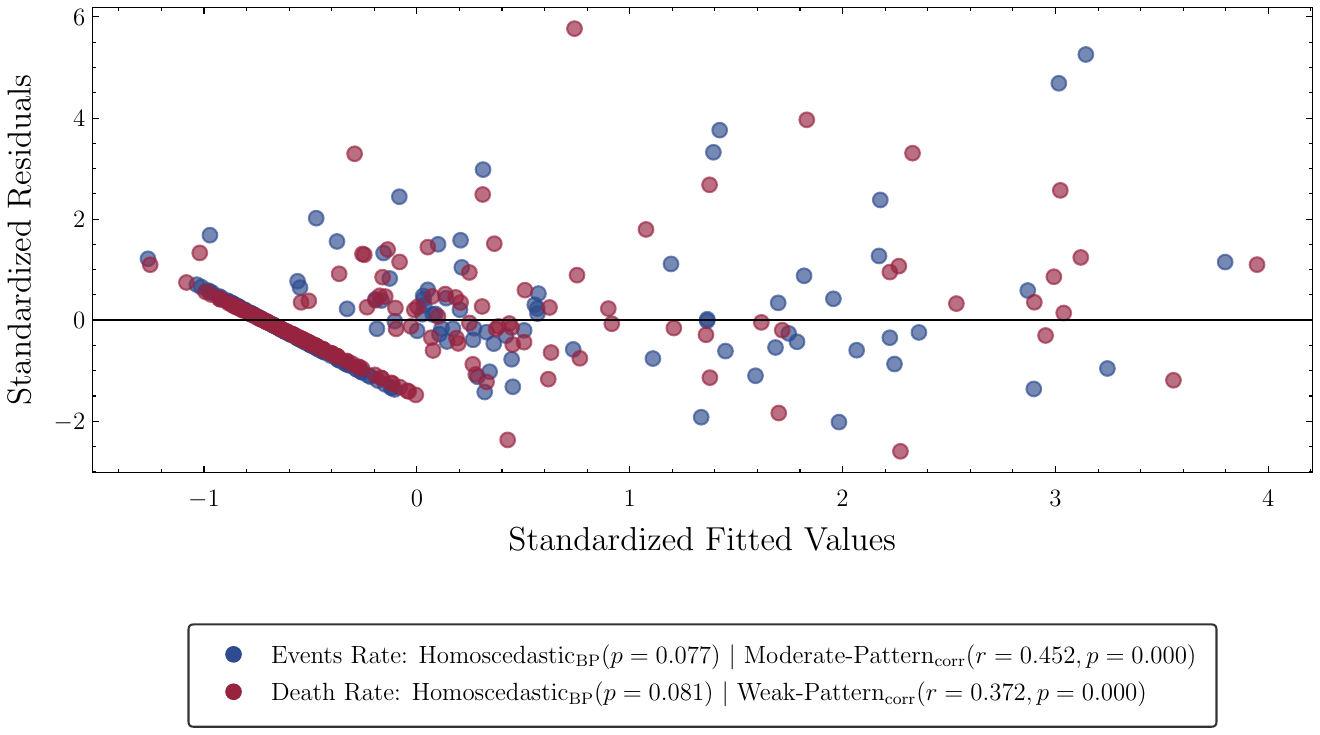}
        \caption{Latin America and The Caribbean}
        \label{heteroLATIN}
    \end{subfigure}
    \hfill
    \begin{subfigure}[b]{0.49\textwidth}
        \includegraphics[width=\textwidth]{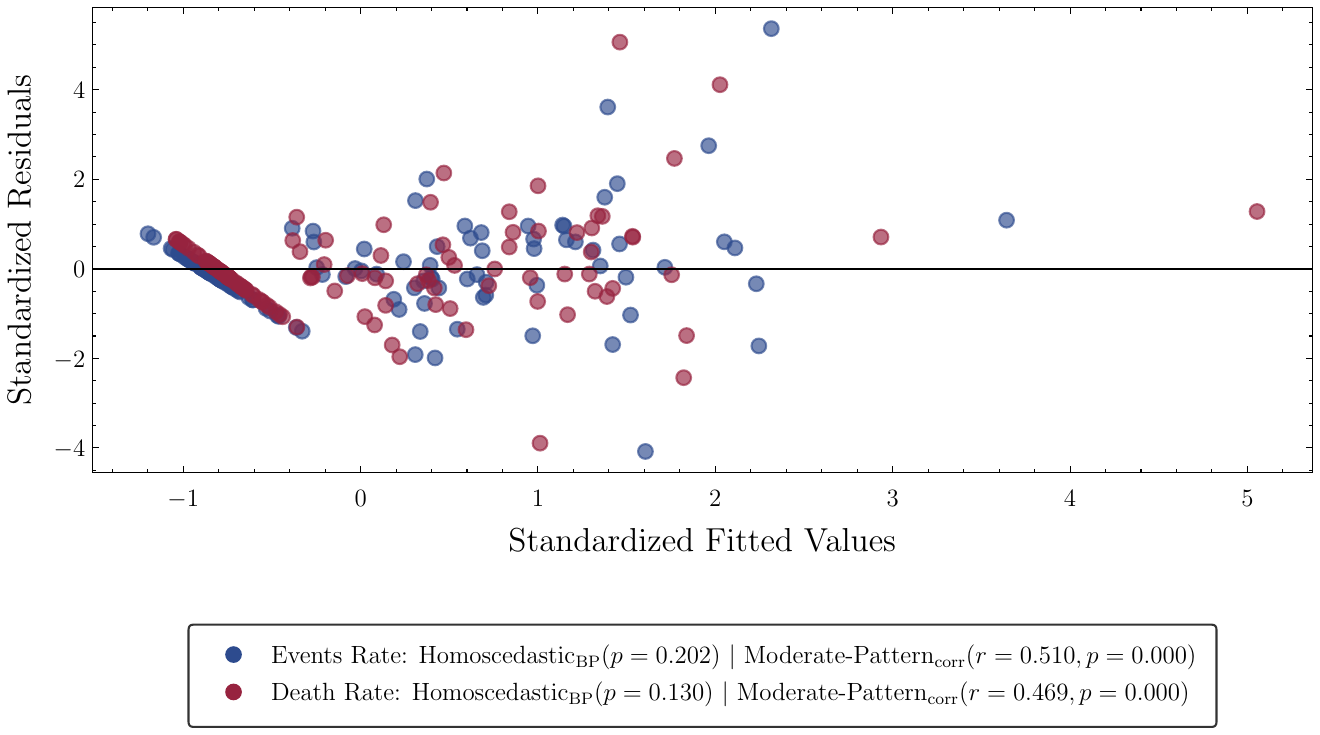}
        \caption{Asia and Pacific}
        \label{figHetero}
    \end{subfigure}
\caption{\textbf{Residuals vs Fitted Values Plot (Homoscedasticity and Linearity Assumption).} Residual plots show Homoscedastic errors across all regions except Sub-Saharan Africa, confirmed by significant p-values from the \citet{79} test. Linearity is assessed via Pearson's Correlation, showing weak to moderate residual correlations. The residuals display erratic patterns with numerous outliers, analyzed in \ref{sectionCook}.}
    \label{resudualsHeteri}
\end{figure}

\noindent We validate the region-wise models in \tabref{tableRegion} by testing Linear Mixed Effects Model (LMEM) assumptions for linearity, normality, and heteroscedasticity \citep{76, 77, 78}. Using the Breusch–Pagan test \citep{79}, only Sub-Saharan Africa models show heteroscedastic residuals for both dependent variables (\figref{heteroSSA}). Other region models demonstrate constant variance in error terms. For linearity assessment, we use the Pearson correlation coefficient \citep{80} to examine residual correlation. A lower correlation coefficient indicates model linearity, as scattered residuals suggest no remaining information to be captured from our variables for predicting the response variable. We observe that residuals are generally spread out, with some clustering patches (\figref{resudualsHeteri}). This is evident in the weak to moderate correlation strength values.

\label{normalityDiscuss}
For normality of residuals, we visually inspect the Quantile-Quantile (QQ) Probability plot of estimated residuals against theoretical residuals (normal distribution) (\appendixref{AppNormal}). A slight S-shaped curve is observed in the plot, with heavy deviations from the quantile lines at both ends, resulting from a likely skewed distribution of conflict rate towards 0. We use two tests to statistically verify the assumption of normality: the Shapiro-Wilk test \citep{81} and the D'Agostino test \citep{82}. Six out of 8 models show normal distribution of residuals through these tests. Two out of the 16 tests conducted show non-normality (Shapiro-Wilk). The two true negatives are not of concern as non-normality doesn't invalidate results in large heterogeneous sample sizes \citep{83}. Additional tests of multicollinearity and data stationarity are carried out and demonstrate model and data robustness (\appendixref{robustnessApp}).

\subsection{Outlier Identification with Cook's Distance}
\label{sectionCook}
\begin{figure}[t]
    \centering
    \begin{subfigure}[b]{0.47\textwidth}
        \includegraphics[width=\textwidth]{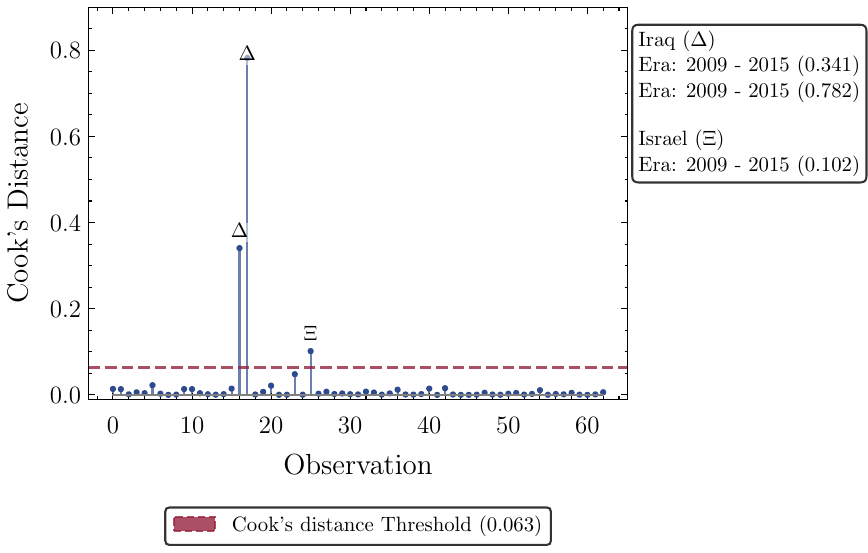}
        \caption{Middle East and North Africa (MENA)}
        \label{cooksDRMENA}
    \end{subfigure}
    \hfill
    \begin{subfigure}[b]{0.47\textwidth}
        \includegraphics[width=\textwidth]{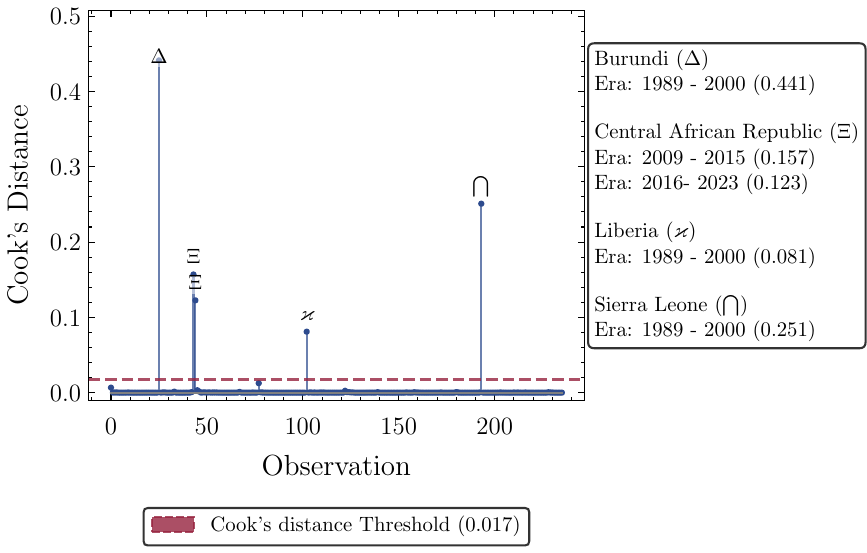}
        \caption{Sub-Saharan Africa}
        \label{cooksDRSubSahara}
    \end{subfigure}
    \vskip\baselineskip
    \begin{subfigure}[b]{0.47\textwidth}
        \includegraphics[width=\textwidth]{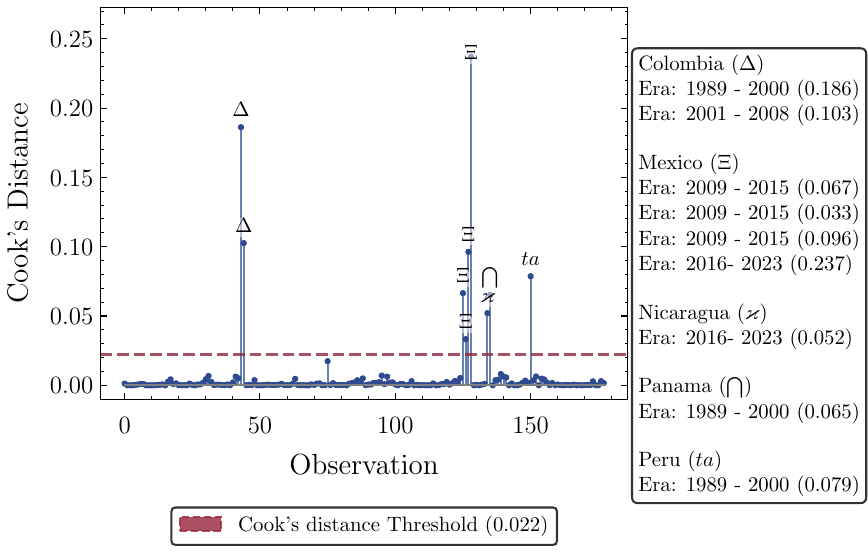}
        \caption{Latin America and The Caribbean}
        \label{cooksDRLatin}
    \end{subfigure}
    \hfill
    \begin{subfigure}[b]{0.47\textwidth}
        \includegraphics[width=\textwidth]{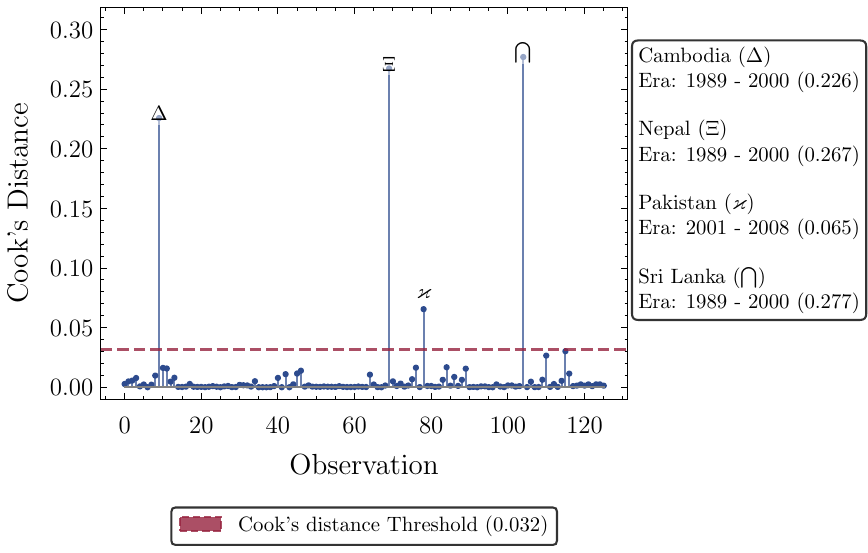}
        \caption{Asia and Pacific}
        \label{cooksDRAsia}
    \end{subfigure}
\caption{\textbf{Outliers in Conflict Death Rate Models with Cook's Distance.} 
The outlier points identified through Cook's D align with periods of major regional conflicts. MENA and Sub-Saharan Africa show higher spikes ($>0.4$), while Latin America/Caribbean and Asia/Pacific have more moderate influential observations ($<0.25$). During extreme events (like Sri Lankan Civil War, Israel-Palestine), the ``typical" polarization-death relationship differs significantly.(\textit{conflict\textsubscript{event-rate}} Outlier Analysis: \appendixref{cooksAppendix})}
    \label{cooksDR}
\end{figure}

\noindent Outliers in socio-economic and political studies often represent meaningful heterogeneity caused by inequalities and rare events \citep{84, 85}, contributing valuable insights to theory building \citep{86}. We use Cook's Distance (D) \citep{87} to identify influential observations by measuring how fitted values change when data points are excluded from a regression. For observation $i$ (removed) from the model, its Cook's D is given as
\begin{equation}
D_i = \frac{\sum_{j=1}^{n} (\hat{y}j - \hat{y}{j(i)})^2}{p*s^2}
\end{equation}
where $\hat{y}\textsubscript{$j$}$ represents the fitted value for observation j, $\hat{y}\textsubscript{$j(i)$}$ is the fitted value for observation j when observation i is removed from the model, p denotes the number of parameters in the regression model, and $s^2$ is the mean squared error. \figref{cooksDR} shows the significant outliers identified through Cook's D for \textit{conflict\textsubscript{death-rate}}.

\section{Discussion}

\subsection{Principal Findings and Regional Outcomes}
\noindent The findings of our study strongly support both initial hypotheses. We observe that various political polarization and fragmentation variables significantly contribute to conflict intensity and severity. The second hypothesis also holds, with region, country and time periods significantly impacting conflict rates, although individually neither country nor time period can predict conflict rates (\figref{icc}).

Region-wise models reveal strong links between significant conflict factors and normative observed geographic differences. In MENA and Sub-Saharan Africa, which are mostly Muslim-majority regions, sectarian violence often occurs between Shia and Sunni ethnic groups - two branches of Islam following different practices \citep{d1,d1a}, this is reflected in our findings. Both regions' models show significant effects of religious principles polarization and religious freedom indices on conflict frequency. Intriguingly, the Sub-Saharan Africa model shows only \textit{Average Term Population} affecting conflict-related deaths. This, coupled with the stark difference between its $R^2_{conditional}$ and $R^2_{marginal}$ values, indicates unmodelled country-specific factors within the region, such as the region's youth bulge \citep{d2}. It's also the only regional model showing heteroskedasticity.

The Latin America and Caribbean model captures conflicts driven by \textit{revolutionary polarizations} and \textit{fundamental social ideologies}, while the Asia-Pacific model reveals more nuanced influences through gender equality and \textit{anti-elitism polarization}. High-income Global South countries like Bahrain, Dominican Republic, and Japan contribute to the negative coefficients for GDP per capita and Gini Index across regional models. This can be seen to have two implications. There are wealth and development disparities within the Global South. Secondly, with era and countries as significant random effects, we can conjecture that economic advancement and inequality reduction has led to decrease of armed conflicts in the region.

\subsection{Parallels with Literature, Normative Ground Truth and its Implications}

\noindent Minority representation emerges as a critical theme in our models. The Global South comprises the marginalized Yazidis and Kurds in the Middle East \citep{d14}, Afro-descendants and indigenous groups in LAC \citep{d15, d15b} and the Rohingyas in South Asia \citep{d16} to name a few. These groups have faced persecution, and have little to no representation in political echelons, while other groups have low tolerance for these groups, leading to \textit{minority identity polarization}. We observe in our models that this leads to higher conflict frequencies in LAC and Asia Pacific, highlighting a glaring issue of marginalization.

Previous studies have focused primarily on affective polarization and political ideology differences in Western countries \citep{d3,d4}, along with how attitudes shape post conflicts \citep{d5}. Similarly, public opinion measurement on armed conflicts is also more straightforward in developed economies \citep{d6}. 
This study's establishing of the strong relationship between armed conflicts and inter-party polarization, should be considered as a focal point for creating more geopolitical harmony and reducing organized violence globally. 

Western perspectives often attribute conflicts in high prone areas of the Global South to lack of education \citep{d7} (Africa) or religious extremism \citep{d8} (MENA), our results reveal a different picture. Higher polarization on issues like political clientelism (vote-buying), political pluralism (co-existence of different groups), and anti-elitism correlates with lower conflict intensities. This suggests that populations in the Global South are politically aware and actively resist problematic leadership - representing the necessary and beneficial side to societal polarization \citep{d9}. This shift parallels broader changes in conflict dynamics, as the Global South moves from international peacekeeping to ad hoc coalitions with regional partners \citep{d10}.

\subsection{Limitations and Future Direction}

\noindent Our study has a few key limitations. The use of polarization and party ideological disparities as proxies for public opinion may miss true sentiment, especially in autocratic systems where political powers may ignore public views. More importantly, macro-level indicators don't necessarily capture individual grievances \citep{d11, d12}, which often directly spark armed conflicts. Similarly, statistical models and p-values have been proven unreliable in their predictive power for civil war and unrest \citep{d13}. Additionally, while V-party \citep{27} provides comprehensive and widely used data, their expert annotations may contain confirmation biases. Despite these limitations, our findings do emphasize the importance of studying political ideologies to reduce fragmentation and ultimately violence in the Global South.


\section{Conclusions and Suggested Urgency}
\noindent This paper examines whether heterogeneity and polarization between ideological stances of political parties impact armed-conflict frequency and deaths in the Global South. We observe that polarization variables affect different regions with varying intensities. Polarization trends adhere to current developments, cultural divisions, and historical predicaments across regions. High polarization of political stances on religious principles and minority rights leads to higher conflict events and deaths across regions. Sub-Saharan Africa shows distinct significant variables with only religion and immigration-based polarization having a relationship with the response variables, indicating peripheral conflict-specific factors in the region.

Our study's findings and their inferences are consequential. The Global South has a higher likelihood of misinformation and disinformation being spread amongst its populace \citep{35}. Additionally, we see evidence of polarization of stances that foster hate and prejudice amongst groups (e.g., Minority, Freedom of Expression, Cultural Superiority). The implications have been seen in Myanmar, where citizens were fed lies and instigated against ethnic minorities on Facebook, leading to mass ethnic cleansing \citep{c1, c2}. It is imperative that alongside focusing on grassroots problems like poverty \citep{10} and horizontal inequalities \citep{11}, key stakeholders create frameworks and policies that reduce polarization propagated by political parties.



\begingroup
\fontsize{9}{11}\selectfont
\newpage
\bibliographystyle{apacite}
\bibliography{main}


\newpage

\appendix

\noindent \huge{\textbf{Appendix}}
\section{Data Description and Statistics}
\label{tableVars}

\begin{table}[htbp]
\renewcommand{\arraystretch}{1.5}
\caption{\textbf{Polarization Variables - Descriptions and Measurement Scales \citep{54}.} Independent Variables of interest in our study cover distinctive political party stances and identities, within a country's election. These variables are used with the original scale and party seat share to compute Daltor Readjusted Polarization Index, for each stance.}
\begin{tabular}{>{\raggedright\arraybackslash}p{3.5cm}p{6cm}p{3.5cm}}
\hline
\textbf{Variable} & \textbf{Description} & \textbf{Original Scale} \\
\hline
\textbf{Anti-Elitism} & Captures polarization in party identities regarding elite groups, where parties stake opposing positions on elite legitimacy & 0 (never anti-elite) to 4 (consistently anti-elite) \\
\textbf{People-Centrism} & Measures polarization in how parties identify with ``the people", contrasting populist stances that claim to represent a unified populace versus pluralistic positions & 0 (never glorifies) to 4 (always glorifies) \\
\textbf{Political Opponents} & Quantifies polarization in party stances toward political opposition, from legitimizing to delegitimizing identities & 0 (always attacks) to 4 (never attacks) \\
\textbf{Political Pluralism} & Measures polarization in party stances on democratic principles and institutions & 0 (not committed) to 4 (fully committed) \\
\textbf{Minority Rights} & Reflects polarization in party stances on minority protection versus majority rule & 0 (always majority) to 4 (never override) \\
\textbf{Rejection of Political Violence} & Captures polarization in party stances on political violence as a legitimate tool & 0 (encourages) to 4 (rejects) \\
\textbf{Immigration} & Measures polarization in party stances on immigration policy & 0 (strongly opposes) to 4 (strongly supports) \\
\textbf{LGBT Social Equality} & Reflects polarization in party stances on LGBT rights and inclusion & 0 (strongly opposes) to 4 (strongly supports) \\
\textbf{Cultural Superiority} & Quantifies polarization in party stances on cultural/national supremacy & 0 (strongly promotes) to 4 (strongly opposes) \\
\textbf{Religious Principles }& Measures polarization in party stances on religion's role in politics & 0 (always religious) to 4 (never religious) \\
\textbf{Working Women} & Captures polarization in party stances on women's economic participation & 0 (strongly opposes) to 4 (strongly supports) \\
\textbf{Economic Left-Right} & Reflects polarization in party stances on economic intervention & 0 (far-left) to 6 (far-right) \\
\textbf{Welfare} & Measures polarization in party stances on welfare policy design & 0 (opposes all) to 5 (universal only) \\
\textbf{Clientelism} & Captures polarization in party stances on vote-buying practices & 0 (no clientelism) to 4 (main strategy) \\
\hline
\multicolumn{3}{l}{\footnotesize Note: Variable descriptions and scales are based on the V-Party Dataset \citep{54}.} \\
\end{tabular}
\end{table}

\begin{figure}[htbp]
    \centering
    \begin{subfigure}[b]{0.48\textwidth}
        \centering
        \includegraphics[width=\textwidth]{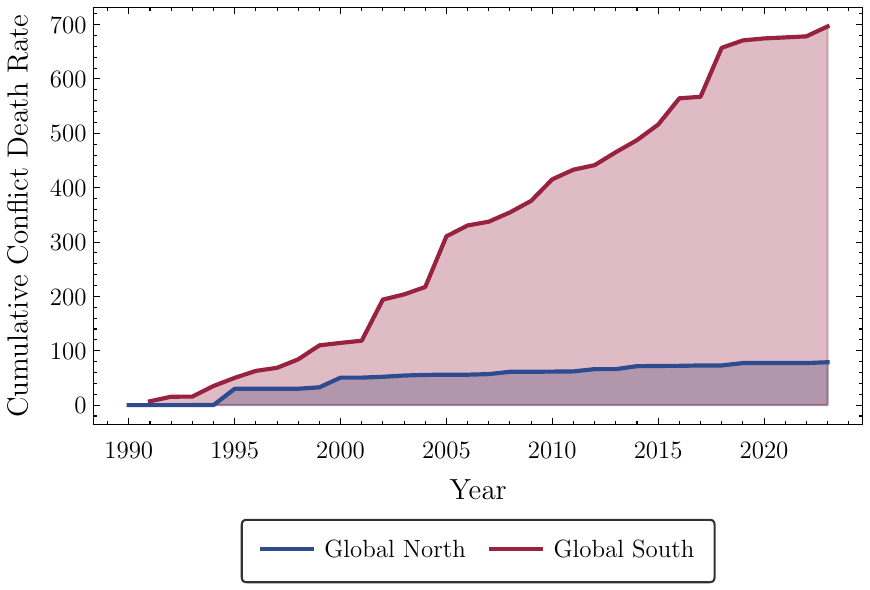}
        \caption{Conflict Death Rate}
        \label{fig:death_rate}
    \end{subfigure}
    \hfill
    \begin{subfigure}[b]{0.48\textwidth}
        \centering
        \includegraphics[width=\textwidth]{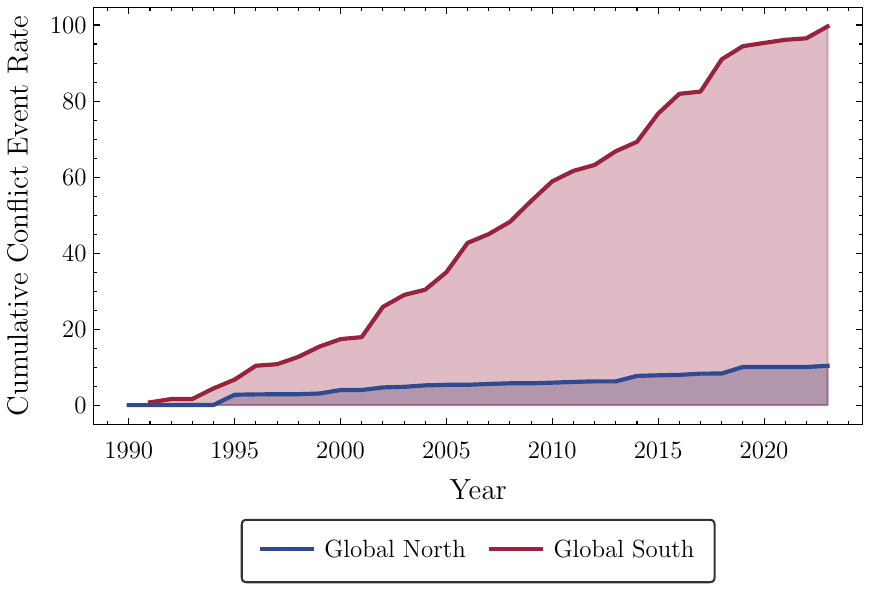}
        \caption{Conflict Event Rate}
        \label{fig:event_rate}
    \end{subfigure}
    \caption{\textbf{Cumulative Conflict Trends in the Global North and South (1990-2020).} The graphs show stark disparities between regions, with the Global South experiencing significantly higher rates in both metrics. \textbf{(a)} The cumulative conflict death rate shows a particularly sharp increase in the Global South post-2005, while the Global North maintains relatively lower levels . \textbf{(b)} Similarly, the conflict event rate demonstrates a consistent upward trend in the Global South, especially accelerating after 2000, while the Global North shows minimal increase. The data in the figures are from UCDP \citep{52}, 2024 version.}
    \label{appendixConflictTrends}
\end{figure}
\label{Appendixconflictrend}

\begin{table}[H]
\renewcommand{\arraystretch}{1.5}
\caption{\textbf{Regional-wise Political Polarization and Conflict Intensity Data Description}}
\centering
\setlength{\tabcolsep}{6pt}
\begin{tabular}{p{2.8cm}cccc}
\hline
& MENA & Sub-Saharan & Latin America & Asia \& Pacific \\
\hline
\textbf{Characteristics} & & & & \\
Unique Countries & 11 & 42 & 22 & 19 \\
Total Observations (N) & 63 & 236 & 178 & 127 \\
Conflict Death Rate Mean ($\bar{x}$) & 2.764 & 1.388 & 0.637 & 0.490 \\
Conflict Event Rate Mean ($\bar{x}$) & 0.611 & 0.119 & 0.104 & 0.097 \\
\hline
\textbf{Era Distribution (\%)} & & & & \\
1989 - 2000 & 30.2 & 36.0 & 41.6 & 38.6 \\
2001 - 2008 & 28.6 & 27.1 & 24.7 & 26.0 \\
2009 - 2015 & 28.6 & 23.3 & 23.0 & 22.0 \\
2016 - 2023 & 12.7 & 13.6 & 10.7 & 13.4 \\
\hline
\textbf{Polarization Variables} & & & & \\
Highest Variance & Anti-Elitism, & Anti-Elitism, & Anti-Elitism, & Cultural Superiority, \\
& Clientelism & Minority Rights & Clientelism & Religious Principles \\
Lowest Variance & Working Women, & People-Centrism, & Working Women, & Immigration, \\
& Immigration & Economic Left-Right & Gender Equality & Gender Equality \\
\hline
\multicolumn{5}{l}{\footnotesize Note: Values for conflict rates represent means per 100,000 population across all countries and time periods.} \\
\multicolumn{5}{l}{\footnotesize Era distribution shows proportion of observations in each time period.}
\end{tabular}
\end{table}

\begin{figure}[htbp]
    \centering
    \begin{subfigure}[b]{0.48\textwidth}
        \centering
        \includegraphics[width=\textwidth]{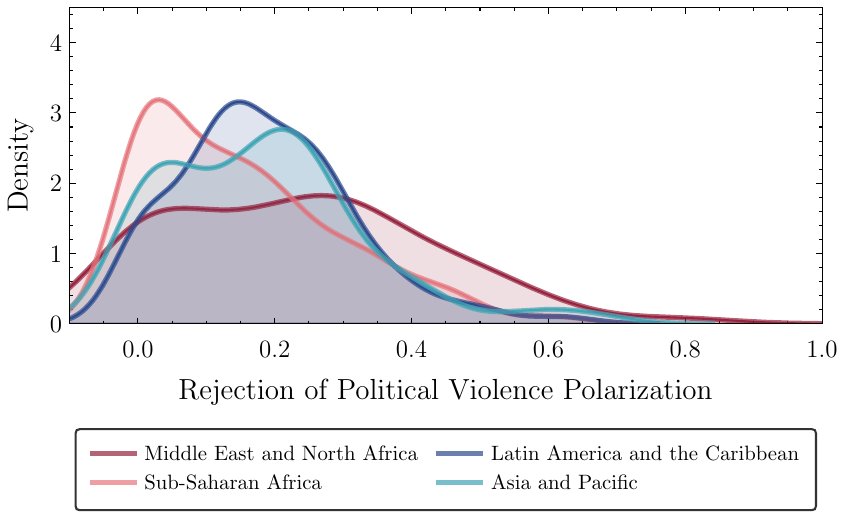}
        \caption{Rejection of Political Violence}
        \label{fig:pol_violence}
    \end{subfigure}
    \hfill
    \begin{subfigure}[b]{0.48\textwidth}
        \centering
        \includegraphics[width=\textwidth]{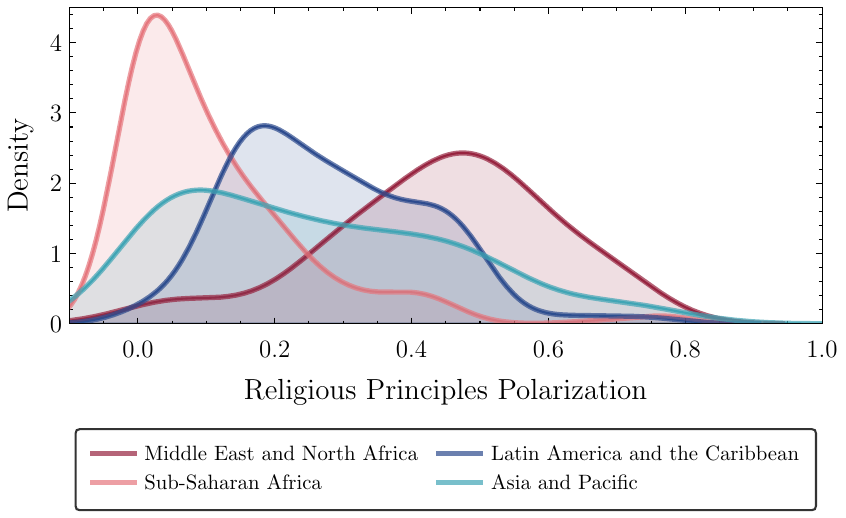}
        \caption{Religious Principles}
        \label{fig:rel_principles}
    \end{subfigure}
    \caption{\textbf{Distribution of key Polarization Variables across Regions.} These variables show strongest significance in our models (\tabref{tableRegion}). \textbf{(a)} Rejection of Political Violence exhibits high variance in MENA and Asia-Pacific, aligning with their significant negative and positive coefficients respectively. \textbf{(b)} Religious Principles shows notable polarization in MENA and Sub-Saharan Africa, consistent with their strong positive associations in the models.}
    \label{fig:polarization_dist}
\end{figure}

\section{Model Results}
\label{modelresultsAppendix}

\begin{table}[H]
    \caption{\textbf{Region-wise Modelling: Linear Mixed Effects Regression Results with Coefficients.} Coefficients are reported with significance levels indicated by asterisks. The detailed interpretations are discussed in the main findings. ER and DR refer to the \textit{conflict\textsubscript{$event_rate$}} and \textit{conflict\textsubscript{$death_rate$}} respectively.}
    \label{tableRegionFull}
    \centering
    \setlength{\tabcolsep}{4pt}  
    \begin{tabular}{l *{4}{cc}}
    \toprule[1.5pt]
    & \multicolumn{2}{c}{MENA} & \multicolumn{2}{c}{Sub-Saharan} & \multicolumn{2}{c}{Latin America} & \multicolumn{2}{c}{Asia \& Pacific} \\
    \cmidrule(lr){2-3} \cmidrule(lr){4-5} \cmidrule(lr){6-7} \cmidrule(lr){8-9}
    & ER & DR & ER & DR & ER & DR & ER & DR \\
    \midrule
    \multicolumn{9}{l}{\textbf{Polarization Variables}} \\
    Anti-Elitism & -0.91 & 0.24 & 0.14 & 0.01 & -0.09 & -0.76 & -0.24 & -1.83*** \\
    People-Centrism & 1.45 & 0.27 & 0.04 & -1.65 & -0.26 & -0.66 & -0.31 & 0.45 \\
    Political Opponents & -1.00 & -1.50 & -0.22 & 1.00 & 0.01 & -1.30 & 0.23 & 0.50 \\
    Political Pluralism & -0.97** & -2.91 & -0.20 & -2.52 & -1.21*** & -4.61*** & -0.11 & -0.11 \\
    Minority Rights & 1.42 & 12.40* & 0.11 & 0.69 & 0.92*** & 3.75 & 0.45*** & 0.40 \\
    Rejection of Political Violence & -2.69* & -5.28 & -0.27 & -0.22 & -1.04*** & -3.84*** & 0.26* & 1.71** \\
    Immigration & 0.30 & 4.07 & -0.54* & -1.02 & -0.23 & 1.30 & -0.05 & 0.54 \\
    LGBT Social Equality & -1.09 & -16.21*** & -0.12 & -0.89 & 0.02 & -1.06 & 0.03 & 0.49 \\
    Cultural Superiority & 2.82* & 2.37 & 0.46 & 0.11 & 0.46* & -0.53 & -0.25 & -0.81 \\
    Religious Principles & 1.54** & 15.14*** & 0.37** & 2.38* & 0.15 & -0.87 & 0.12 & 0.83 \\
    Gender Equality & -0.67 & 1.48 & 0.03 & 0.09 & 0.23 & 1.92 & -0.37*** & -2.17** \\
    Working Women & 0.81 & 3.46 & 0.30 & 2.47 & 0.00 & 1.88 & 0.09 & -0.15 \\
    Economic Left-Right & 2.94*** & 7.90 & -0.17 & -0.39 & 0.07 & 1.76 & -0.17 & 0.27 \\
    Welfare & -0.56** & -6.00 & 0.25 & -1.59 & 0.39* & 0.39 & 0.10 & -0.54 \\
    Clientelism & -0.45 & -4.24 & 0.01 & 0.82 & -0.43** & -0.39 & 0.30 & 1.29 \\
    \midrule
    \multicolumn{9}{l}{\textbf{Control Variables}} \\
    Avg. Term Population & -0.00** & -0.00 & 0.00*** & 0.00** & 0.00*** & 0.00*** & -0.00 & -0.00 \\
    Period Length & -0.06 & -0.56 & 0.01 & 0.02 & -0.02 & -0.07 & 0.02*** & 0.08** \\
    Freedom Expression & -2.53** & -12.30 & -0.54** & -1.34 & -1.16*** & -6.38*** & 0.15 & 0.79 \\
    Religious Freedom & 0.65 & 3.06*** & -0.01 & -0.49 & -0.07 & -1.20*** & 0.01 & -0.05 \\
    GDP per Capita & -0.00** & -0.00*** & -0.00 & -0.00 & -0.00 & 0.00 & -0.00 & -0.00*** \\
    Gini Index & -0.46 & -4.79 & -0.62** & 4.02 & 0.00 & -0.01 & -0.00 & -0.01 \\
    \bottomrule[1.5pt]
    \addlinespace[1ex]
    \multicolumn{9}{l}{\small\textit{Note:} * p $<$ 0.10, ** p $<$ 0.05, *** p $<$ 0.01} \\
    \end{tabular}
\end{table}

\endgroup
\begingroup
\fontsize{9}{11}\selectfont

\section{Model Robustness \& Assumptions Check}

\label{robustnessApp}

\subsection{Normality of Residual (Visualisation)}

\noindent The results of testing of the normality assumption of the model (\figref{normalPLot}), are discussed in depth in the main findings (Section \ref{normalityDiscuss}). The Shapiro-Wilk test used is given as

\begin{equation}
W = \frac{(\sum_{i=1}^n a_i x_{(i)})^2}{\sum_{i=1}^n (x_i - \bar{x})^2}
\end{equation}
where $x_{(i)}$ are the ordered sample values, $a_i$ are the weights derived from the mean, variance and covariance of order statistics of a sample from a normal distribution, n is the sample size, and $\bar{x}$ is the sample mean.

\label{AppNormal}
\begin{figure}[H]
    \centering
    \begin{subfigure}[b]{0.49\textwidth}
        \includegraphics[width=\textwidth]{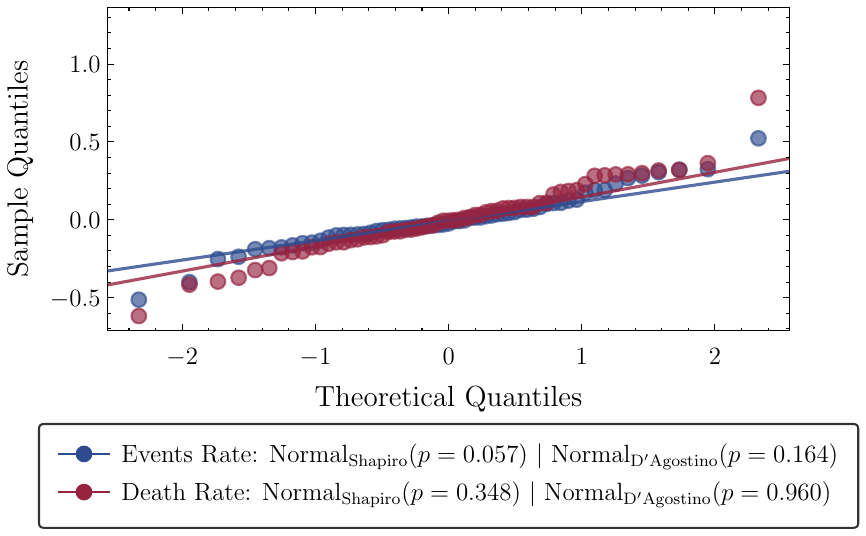}
        \caption{Middle East and North Africa}
        \label{fig:plot1}
    \end{subfigure}
    \hfill
    \begin{subfigure}[b]{0.49\textwidth}
        \includegraphics[width=\textwidth]{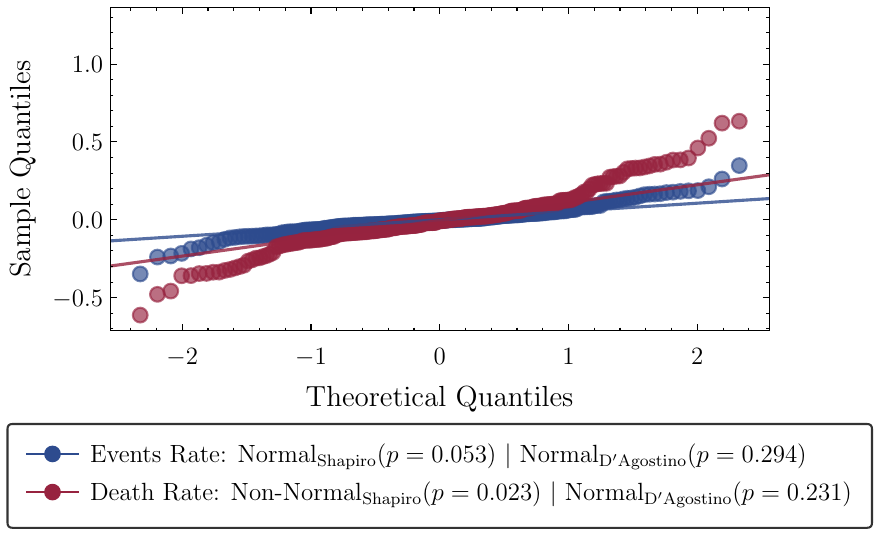}
        \caption{Sub-Saharan Africa}
        \label{fig:plot2}
    \end{subfigure}
    \vskip\baselineskip
    \begin{subfigure}[b]{0.49\textwidth}
        \includegraphics[width=\textwidth]{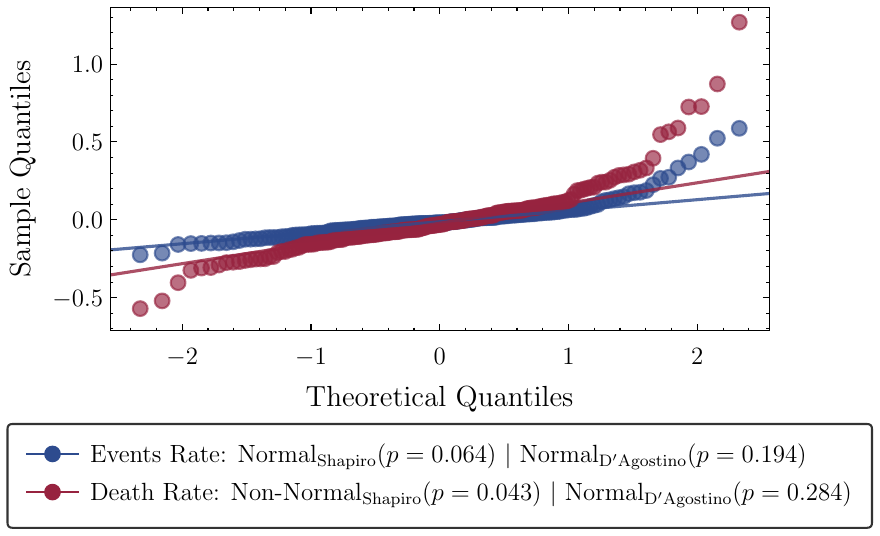}
        \caption{Latin America and the Caribbean}
        \label{fig:plot3}
    \end{subfigure}
    \hfill
    \begin{subfigure}[b]{0.49\textwidth}
        \includegraphics[width=\textwidth]{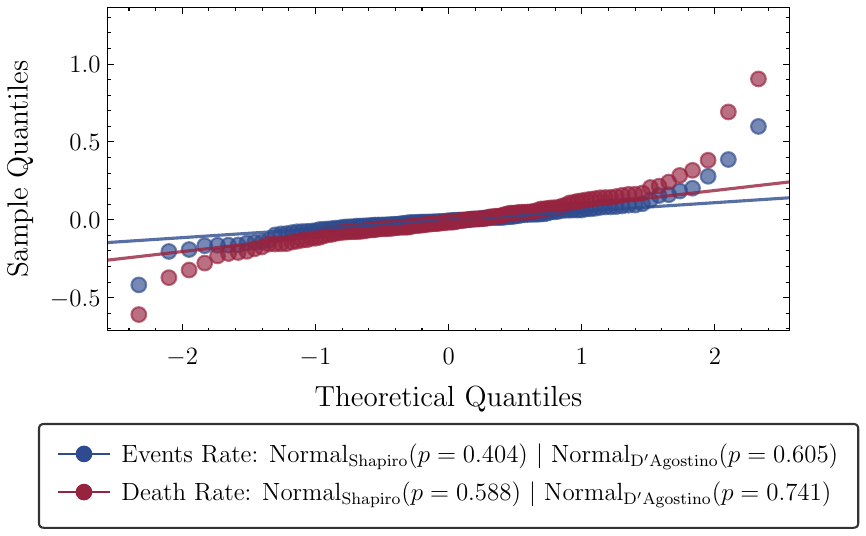}
        \caption{Asia and Pacific}
        \label{figNormality}
    \end{subfigure}
    \caption{\textbf{Q-Q Residuals Plot of Sample (Estimated) Quantiles and Theoretical Quantiles.} Comparing the distribution of residuals in regional models against theoretical normal distributions of the residuals, skewness and heavy left-tailed data distribution is observed across all regions, with S-shaped patterns. This is possibly due to the data being zero-inflated, with many political terms of countries facing zero death rates and frequency of conflicts. The normality assumption holds, though, through the statistical significance tests.}
    \label{normalPLot}
\end{figure}

\endgroup
\begingroup
\fontsize{9}{11}\selectfont

\newpage

\subsection{Multicollinearity}

\noindent Multicollinearity occurs when two or more independent variable pairs are correlated and linearly dependent with each other. The Variance Inflation Factor (VIF) is a measure of multicollinearity. It is the ratio between the variance of a fully-fitted model and the variance of the coefficient being tested. It is given as
\begin{equation}
\text{VIF}_j = \frac{\text{var}(\hat{\beta}j)}{\text{var}(\hat{\beta}j^{\text{single}})} = \frac{\text{MSE\textsubscript{full}}{\text{}}}{\text{MSE\textsubscript{single}}{\text{}}} \cdot \frac{1}{1-R^2_j}
\end{equation}
where $\text{var}(\hat{\beta}_j)$ is the variance of the j-th coefficient in the full model containing all predictors (polarization and control variables), and $\text{var}(\hat{\beta}_j^{\text{single}})$ is the variance from a model with only the j-th predictor.
Upon calculating VIF for our model coefficients \tabref{tableVIF}, we find that all coefficients are below the generally acceptable critical level (VIF$<$10) \citep{A1}. A potential problem can arise with MENA coefficients as they have relatively high VIF scores, with 11 out of the 20 variables having a VIF above 5.

\begin{table}[H]
    \centering
    \setlength{\tabcolsep}{4pt}
    \begin{tabular}{l *{4}{cc}}
    \toprule[1.5pt]
    & \multicolumn{2}{c}{MENA} & \multicolumn{2}{c}{Sub-Saharan} & \multicolumn{2}{c}{Latin America} & \multicolumn{2}{c}{Asia \& Pacific} \\
    \cmidrule(lr){2-3} \cmidrule(lr){4-5} \cmidrule(lr){6-7} \cmidrule(lr){8-9}
    & ER & DR & ER & DR & ER & DR & ER & DR \\
    \midrule
    \multicolumn{9}{l}{\textbf{Polarization Variables}} \\
    Anti-Elitism & 3.50 & 3.50 & 1.69 & 1.69 & 2.28 & 2.28 & 3.51 & 3.51 \\
    People-Centrism & 4.66 & 4.66 & 1.80 & 1.80 & 2.04 & 2.04 & 2.99 & 2.99 \\
    Political Opponents & 5.51 & 5.51 & 2.57 & 2.57 & 2.57 & 2.57 & 2.14 & 2.14 \\
    Political Pluralism & 7.33 & 7.33 & 4.05 & 4.05 & 3.60 & 3.60 & 3.98 & 3.98 \\
    Minority Rights & 4.43 & 4.43 & 1.68 & 1.68 & 2.39 & 2.39 & 3.29 & 3.29 \\
    Rejection of Political Violence & 6.95 & 6.95 & 2.63 & 2.63 & 3.38 & 3.38 & 2.79 & 2.79 \\
    Immigration & 2.59 & 2.59 & 1.89 & 1.89 & 2.17 & 2.17 & 1.83 & 1.83 \\
    LGBT Social Equality & 4.01 & 4.01 & 2.31 & 2.31 & 2.97 & 2.97 & 3.95 & 3.95 \\
    Cultural Superiority & 6.69 & 6.69 & 2.69 & 2.69 & 2.72 & 2.72 & 3.79 & 3.79 \\
    Religious Principles & 2.41 & 2.41 & 2.01 & 2.01 & 2.09 & 2.09 & 4.93 & 4.93 \\
    Gender Equality & 6.20 & 6.20 & 1.77 & 1.77 & 1.86 & 1.86 & 2.52 & 2.52 \\
    Working Women & 3.43 & 3.43 & 1.49 & 1.49 & 2.70 & 2.70 & 2.73 & 2.73 \\
    Economic Left-Right & 4.50 & 4.50 & 1.41 & 1.41 & 3.93 & 3.93 & 2.55 & 2.55 \\
    Welfare & 4.31 & 4.31 & 1.65 & 1.65 & 1.99 & 1.99 & 2.76 & 2.76 \\
    Clientelism & 7.63 & 7.63 & 2.11 & 2.11 & 1.92 & 1.92 & 3.74 & 3.74 \\
    \midrule
    \multicolumn{9}{l}{\textbf{Control Variables}} \\
    Avg. Term Population & 9.78 & 9.78 & 1.35 & 1.35 & 1.72 & 1.72 & 2.91 & 2.91 \\
    Period Length & 1.42 & 1.42 & 1.14 & 1.14 & 1.32 & 1.32 & 1.48 & 1.48 \\
    Freedom Expression & 9.91 & 9.91 & 1.74 & 1.74 & 2.70 & 2.70 & 2.88 & 2.88 \\
    Religious Freedom & 8.38 & 8.38 & 1.32 & 1.32 & 2.29 & 2.29 & 2.57 & 2.57 \\
    GDP per Capita & 9.94 & 9.94 & 1.13 & 1.13 & 1.73 & 1.73 & 1.93 & 1.93 \\
    Gini Index & 3.19 & 3.19 & 1.24 & 1.24 & 1.60 & 1.60 & 1.50 & 1.50 \\
    \bottomrule[1.5pt]
    \end{tabular}
    \caption{\textbf{Variance Inflation Factors (VIF) Across Regions.} This table presents VIF values for both Event Rate (ER) and Death Rate (DR) models across regions. VIF values indicate the severity of multicollinearity, with values below 5 generally considered acceptable, 5-10 potentially concerning, and above 10 indicating severe multicollinearity.}
    \label{tableVIF}
\end{table}

\subsection{Cook's D Outlier Analysis for Conflicts Event Rate}
\label{cooksAppendix}
\begin{figure}[H]
    \centering
    \begin{subfigure}[b]{0.49\textwidth}
        \includegraphics[width=\textwidth]{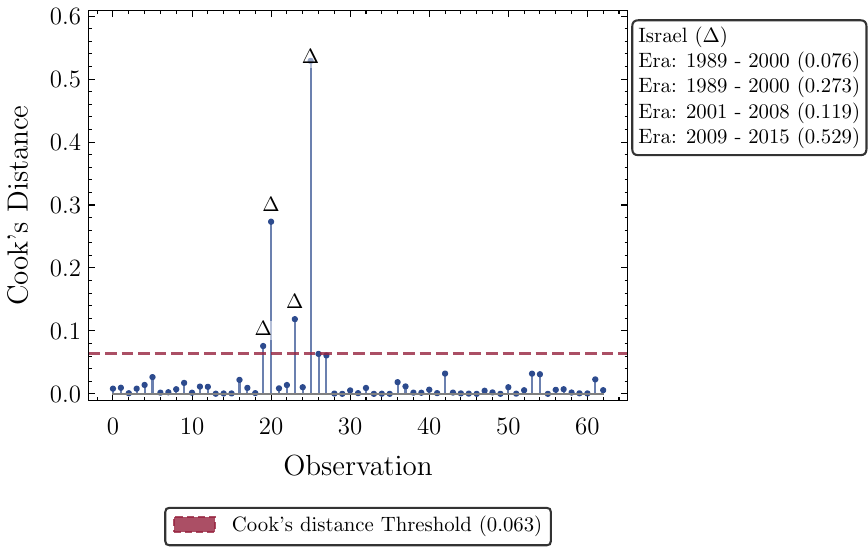}
        \caption{Middle East and North Africa (MENA)}
        \label{cooksERMENA}
    \end{subfigure}
    \hfill
    \begin{subfigure}[b]{0.49\textwidth}
        \includegraphics[width=\textwidth]{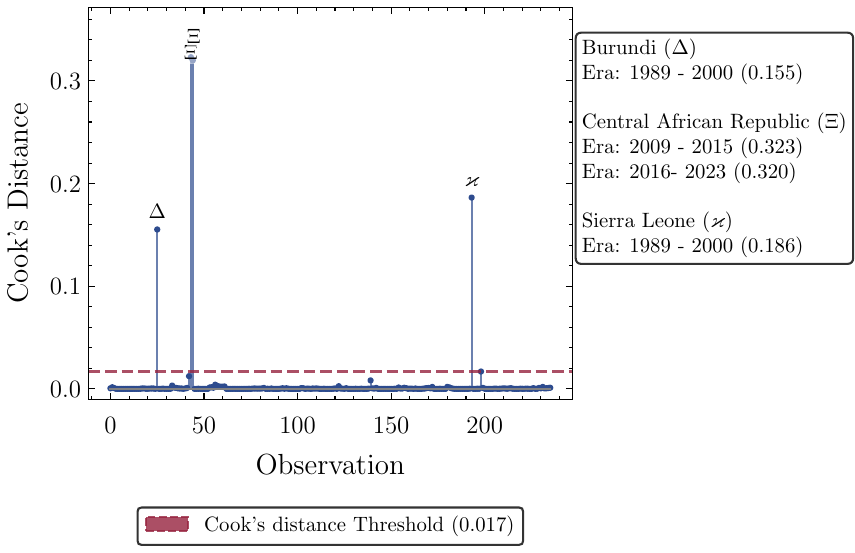}
        \caption{Sub-Saharan Africa}
        \label{cooksERSubSahara}
    \end{subfigure}
    \vskip\baselineskip
    \begin{subfigure}[b]{0.49\textwidth}
        \includegraphics[width=\textwidth]{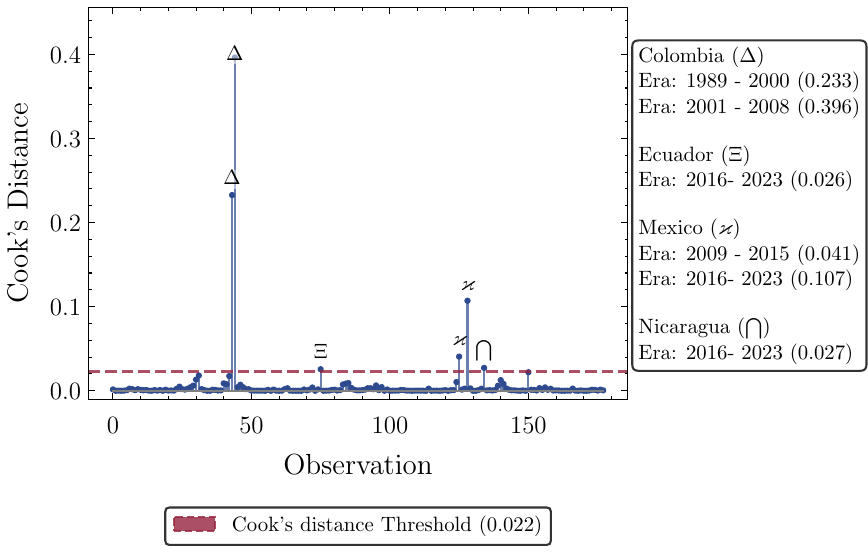}
        \caption{Latin America and The Caribbean}
        \label{cooksERLatin}
    \end{subfigure}
    \hfill
    \begin{subfigure}[b]{0.49\textwidth}
        \includegraphics[width=\textwidth]{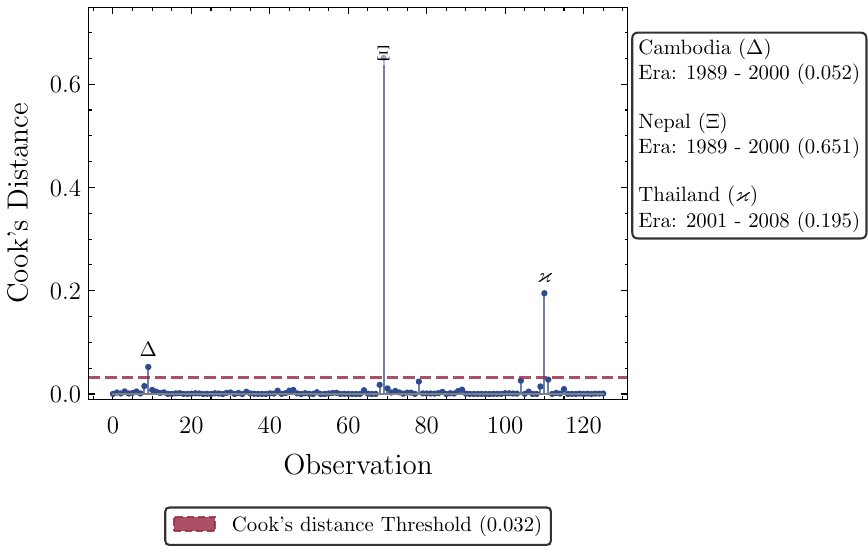}
        \caption{Asia and Pacific}
        \label{cooksERAsia}
    \end{subfigure}
\caption{\textbf{Outliers in Conflict Event Rate Models with Cook's Distance.}
Latin America and The Caribbean model show prominent outliers, particularly with Colombia exhibiting the highest peaks (around 0.4) in the observations. In the MENA model, Israel stands out as the sole but significant outlier country, with distinct peaks from 1989 to 2015. While Asia and Pacific generally shows moderate patterns, Nepal notably emerges as a surprising outlier with the highest peak (approximately 0.65) during the 1989-2000 period (during a Maoist insurgency in the country), surpassing all other regions' outlier magnitudes. When compared to the death rate outlier analysis in \figref{cooksDR}, the distribution and intensity of outliers show distinct patterns, with certain regions having repeating countries.}
    \label{cooksER}
\end{figure}

\subsection{Stationarity of Data}
\noindent While not using a time-series model or forecasting as our core objective, the data used in the study does have a time series element. Hence, we check the data for stationarity using both the Augmented Dickey-Fuller (ADF) test \citep{A3}  and the KPSS test (\citep{A3}). Stationarity indicates that the statistical properties (mean, variance, autocorrelation) of our time-series data remain constant over time. We group our data by region, sort it according to each regime's beginning year, and then perform these tests. As shown in \tabref{tableStationarity}, high stationarity is observed in the data, indicating that time-series forecasting and predictive modelling would be appropriate for this dataset. We suggest robust models like ARIMA, SARIMA and Prophet for this purpose.

\begin{table}[H]
    \centering
    \setlength{\tabcolsep}{4pt}
    \begin{tabular}{l *{4}{cc}}
    \toprule[1.5pt]
    & \multicolumn{2}{c}{MENA} & \multicolumn{2}{c}{Sub-Saharan} & \multicolumn{2}{c}{Latin America} & \multicolumn{2}{c}{Asia \& Pacific} \\
    \cmidrule(lr){2-3} \cmidrule(lr){4-5} \cmidrule(lr){6-7} \cmidrule(lr){8-9}
    & ADF & KPSS & ADF & KPSS & ADF & KPSS & ADF & KPSS \\
    \midrule
    \multicolumn{9}{l}{\textbf{Polarization Variables}} \\
    Anti-Elitism & \checkmark & $\times$ & \checkmark & \checkmark & \checkmark & \checkmark & \checkmark & \checkmark \\
    People-Centrism & \checkmark & \checkmark & \checkmark & \checkmark & \checkmark & \checkmark & \checkmark & \checkmark \\
    Political Opponents & \checkmark & $\times$ & \checkmark & \checkmark & \checkmark & \checkmark & \checkmark & \checkmark \\
    Political Pluralism & \checkmark & \checkmark & \checkmark & \checkmark & \checkmark & \checkmark & \checkmark & \checkmark \\
    Minority Rights & \checkmark & \checkmark & \checkmark & \checkmark & \checkmark & \checkmark & \checkmark & \checkmark \\
    Rejection of Political Violence & \checkmark & $\times$ & \checkmark & \checkmark & \checkmark & \checkmark & \checkmark & \checkmark \\
    Immigration & \checkmark & \checkmark & \checkmark & \checkmark & \checkmark & \checkmark & \checkmark & \checkmark \\
    LGBT Social Equality & \checkmark & \checkmark & \checkmark & \checkmark & \checkmark & \checkmark & \checkmark & \checkmark \\
    Cultural Superiority & \checkmark & \checkmark & \checkmark & \checkmark & \checkmark & \checkmark & \checkmark & \checkmark \\
    Religious Principles & \checkmark & \checkmark & \checkmark & \checkmark & \checkmark & \checkmark & \checkmark & \checkmark \\
    Gender Equality & \checkmark & \checkmark & \checkmark & \checkmark & \checkmark & \checkmark & \checkmark & \checkmark \\
    Working Women & \checkmark & \checkmark & \checkmark & \checkmark & \checkmark & \checkmark & \checkmark & \checkmark \\
    Economic Left-Right & \checkmark & \checkmark & \checkmark & \checkmark & \checkmark & \checkmark & \checkmark & \checkmark \\
    Welfare & \checkmark & \checkmark & \checkmark & \checkmark & \checkmark & \checkmark & \checkmark & \checkmark \\
    Clientelism & \checkmark & \checkmark & \checkmark & \checkmark & \checkmark & \checkmark & \checkmark & \checkmark \\
    \midrule
    \multicolumn{9}{l}{\textbf{Control Variables}} \\
    Avg. Term Population & \checkmark & \checkmark & \checkmark & \checkmark & \checkmark & \checkmark & \checkmark & \checkmark \\
    Period Length & \checkmark & \checkmark & \checkmark & \checkmark & \checkmark & \checkmark & \checkmark & \checkmark \\
    Freedom Expression & \checkmark & \checkmark & \checkmark & \checkmark & \checkmark & \checkmark & \checkmark & $\times$ \\
    Religious Freedom & \checkmark & \checkmark & \checkmark & \checkmark & \checkmark & $\times$ & \checkmark & $\times$ \\
    GDP per Capita & \checkmark & \checkmark & \checkmark & $\times$ & $\times$ & $\times$ & \checkmark & $\times$ \\
    Gini Index & \checkmark & $\times$ & \checkmark & \checkmark & \checkmark & \checkmark & \checkmark & \checkmark \\
    \midrule
    Stationary (\%) & 100\% & 81\% & 100\% & 95.2\% & 95.2\% & 90.5\% & 100\% & 85.7\% \\
     \bottomrule[1.5pt]
    \addlinespace[1ex]
    \multicolumn{9}{l}{\small\textit{Notes:} \checkmark indicates stationarity, $\times$ indicates non-stationarity.} \\
    \end{tabular}
    \caption{\textbf{Stationarity Test Results Across Regions.} This table presents the results of Augmented Dickey-Fuller (ADF) and KPSS tests across regions. Percentages show the proportion of variables that are stationary according to each test.}
    \label{tableStationarity}
\end{table}

\endgroup

\end{document}